\documentclass[]{aa}
\usepackage{graphicx}
\usepackage[varg]{txfonts}

 \usepackage{natbib,twoopt}

 \bibpunct{(}{)}{;}{a}{}{,}             

\usepackage[breaklinks=true]{hyperref} 
\hypersetup{colorlinks=true,urlcolor=blue,citecolor=blue,pdfborder= 0 0 0}

 \defcitealias{2012MNRAS.423.2279C}{C12}
 \defcitealias{2010ApJ...717...74Z}{Z10}
 \defcitealias{2010ApJ...714.1715F}{F10}

\begin{document}

\title{Missing baryons traced by the galaxy luminosity density in the large-scale WHIM filaments}

\author{J. Nevalainen\inst{1}\fnmsep\thanks{jukka@to.ee}
          \and
          E. Tempel\inst{1}
          \and
          L. J. Liivam{\"{a}}gi\inst{1}
          \and      
          E. Branchini\inst{2,3,4}
          \and
          M. Roncarelli\inst{5}
          \and
          C. Giocoli \inst{5,6}
          \and
          P. Hein{\"{a}}m{\"{a}}ki \inst{7}
          \and \\
          E. Saar\inst{1} 
          \and
          A. Tamm\inst{1} 
          \and
          A. Finoguenov\inst{8} 
          \and
          P. Nurmi \inst{7}
          \and
          M. Bonamente\inst{9}}

\date{Accepted Aug 2015}

\institute{Tartu Observatory, Observatooriumi 1, 61602 T{\~{o}}ravere, Estonia
\and
Dipartimento di Matematica e  Fisica, Universit\`a degli Studi ``Roma Tre'', via della Vasca Navale 84, I-00146 Roma, Italy
\and
INFN, Sezione di Roma Tre, via della Vasca Navale 84, I-00146 Roma, Italy
\and
INAF, Osservatorio Astronomico di Roma, Monte Porzio Catone, Italy
\and
University of Bologna,  viale Berti Pichat 6/2, I-40127 Bologna, Italy
\and
Aix Marseille Universit{\'{e}}, CNRS, LAM (Laboratoire d’Astrophysique de Marseille) UMR 7326, 13388 Marseille, France
\and
Tuorla Observatory, V{\"{a}}is{\"{a}}l{\"{a}}ntie 20, FI-21500  Piikki{\"{o}}, Finland
\and
Department of Physics, University of Helsinki, Gustaf H{\"{a}}llstr{\"{o}}min katu 2a, 00014, Helsinki, Finland
\and
University of Alabama in Huntsville, Huntsville, AL 35899, USA
}

\abstract{
We propose a new approach to the missing baryons problem. Building on the common assumption that the missing baryons are in the form of the Warm Hot 
Intergalactic Medium (WHIM), we further assumed here that the galaxy luminosity density can be used as a tracer of the WHIM. The latter assumption is supported by 
our finding of a significant correlation between the WHIM density and the galaxy luminosity density in the hydrodynamical simulations of Cui et~al. (2012). We 
further found that the fraction of the gas mass in the WHIM phase is substantially (by a factor of $\sim$1.6) higher within the large scale galactic filaments, 
i.e. $\sim$70\%, compared to the average in the full simulation volume of $\sim$0.1~Gpc$^3$. The relation between the WHIM overdensity and the galaxy luminosity 
overdensity within the galactic filaments is consistent with linear: $\delta_{\rm whim}~=~0.7~\pm~0.1~\times~\delta_\mathrm{LD}^{0.9 \pm 0.2}$.

We applied our procedure to the line of sight to the blazar H2356-309 and found evidence for the WHIM in correspondence of the Sculptor Wall (z~$\sim$0.03 and 
$\log{N_H}$ = $19.9^{+0.1}_{-0.3}$) and Pisces-Cetus superclusters (z~$\sim$0.06 and $\log{N_H}$ = $19.7^{+0.2}_{-0.3}$), in agreement with the redshifts and column 
densities of the X-ray absorbers identified and studied by Fang et~al. (2010) and Zappacosta et~al. (2010). This agreement indicates that the galaxy luminosity 
density and galactic filaments are reliable signposts for the WHIM and that our method is robust in estimating the WHIM density. 
The signal that we detected cannot originate from the halos of the nearby galaxies since they cannot account for the 
large WHIM column densities that our method and X-ray analysis consistently find in the Sculptor Wall and Pisces-Cetus superclusters.}


\maketitle

\section{Introduction}
\label{intro}

A large fraction of the local ($z<1$) baryons as predicted by the concordance $\Lambda$CDM cosmology are not detected \citep[e.g.][]{2008Sci...319...55N, 2012ApJ...759...23S}, giving rise to the missing baryons problem. Since the current cosmological model agrees very accurately with most of the observations, it is likely that the problem of these missing baryons is observational by nature, rather than a problem with the cosmological theory.

In the current view the cosmological large-scale structure formation is driven by dark matter, due to its mass dominance and solely gravitating nature. First the dark matter skeleton of the cosmic web is formed, and then the baryons fall and condense to these filamentary gravitational potential wells 
(e.g. \citealt{1999ApJ...514....1C}; \citealt{2012MNRAS.423.2279C}, hereafter \citetalias{2012MNRAS.423.2279C}). With cosmological time these dense filamentary regions become more pronounced and finally at $z<1$ the released gravitational potential is sufficient to shock-heat the baryons at over-densities $\delta_{b} = \rho_{b}/\langle\rho_{b}\rangle$ in the range 1--100 to temperatures of $10^{5}$--$10^{7}$~K \citep[e.g.][]{2001ApJ...552..473D, 2007ARA&A..45..221B}. The surrounding medium is also enriched by metals through galactic outflows and stellar feedback \citep[e.g.][]{2006ApJ...650..560C, 2010MNRAS.402.1536S}. The gas in this phase is usually called the Warm Hot Intergalactic Medium (WHIM).

The expected low density of the WHIM renders it very difficult to detect in emission and thus a viable candidate for the (so far) missing baryons. The rare detections of WHIM in emission have been obtained by e.g. \citet{1999A&A...341...23K} and \citet{2005MNRAS.357..929Z} using ROSAT/PSPC observations of the Shapley and Sculptor superclusters, respectively and by \citet{2008A&A...482L..29W} for a cluster pair A222/A223. The full understanding of these detections requires measuring the redshift of the emission, which is currently beyond the capabilities of Chandra and XMM-Newton telescopes. The situation will only improve with advent of Athena mission. The absorption measurements in X-rays (e.g. \citealt{2010ApJ...717...74Z}, hereafter \citetalias{2010ApJ...717...74Z}; \citealt{2010ApJ...714.1715F}, hereafter \citetalias{2010ApJ...714.1715F}; \citealt{2014ApJ...782L...6R}) and especially in far ultra-violet \citep{1998AJ....116.2094S, 2003ApJ...594L.107S, 2012ApJ...759..112T, 2014ApJ...791..128S} have been quite successful. However, they are limited by the small number of suitable background sources observable in the bright state behind significant WHIM structures.

The filamentary regions with enhanced gas density along the cosmic web, in addition to hosting WHIM, are also preferential sites of galaxy formation and stellar feedback process, see \citep[e.g.][]{2013PASA...30...30B, 2012MNRAS.423.2279C, 2006ApJ...650..560C, 2014ApJ...788..157N, 2000ApJ...540...62S}. Thus galaxies and the WHIM are expected to trace the same underlying large-scale filaments and, therefore, should be spatially correlated. Building on this assumption we have developed a method to trace the missing baryons in the form of WHIM using the observed galaxies and their luminosity density as a proxy. The method consists of
1) detecting the galaxy filaments in spectroscopic galaxy catalogues, 
2) determining the galaxy luminosity density fields around the detected filaments,
3) deriving and calibrating a relation between the galaxy luminosity density and the WHIM density from hydrodynamical N-body simulations in cosmological boxes and
4) applying the above phenomenological relation to estimate the WHIM density using the observed luminosity density of galaxies in the filaments.

We tested the accuracy of our WHIM column density estimation by applying it to the 2dF data around the Sculptor Wall (SW) and Pisces-Cetus (PC) superclusters and comparing our values with those obtained via the X-ray absorption measurements for these structures (\citealt{2009ApJ...695.1351B}; \citetalias{2010ApJ...714.1715F}; \citetalias{2010ApJ...717...74Z}).

Our definition of WHIM is the gas in the temperature range of $10^{5}$--$10^{7}$~K and in the baryon overdensity range $\delta_\mathrm{b} = \rho_\mathrm{b}/\langle\rho_\mathrm{b}\rangle$ of 1--100. We use $\Omega_{\rm m} = 0.3$, $\Omega_{\Lambda} = 0.7$ and $H_0 = 70~\mathrm{km~s}^{-1}\mathrm{Mpc}^{-1}$.

\section{Methods}
\subsection{Filament detection}
\label{bisous}

The detection of filaments is performed by applying an object point process with interactions i.e. the Bisous process \citep{Stoica:05} to the spatial three-dimensional distribution of galaxies. The method provides a quantitative classification that agrees well with a visual impression of the filamentary nature of the cosmic web and is based on a robust and well-defined mathematical scheme. More details regarding the Bisous model can be found in \citet{2007JRSSC..56....1S, 2010A&A...510A..38S} and \citet{2014MNRAS.438.3465T}. For reader convenience, a brief summary is provided below.

The Bisous model approximates the filamentary web by a random configuration of small cylindrical segments. The model assumes that locally galaxy distribution can be probed with relatively small cylinders, which can be combined to trace a galaxy filament if the neighboring cylinders are oriented similarly. One of the advantages of such approach is that it relies directly on the observed positions of galaxies and does not require any additional smoothing kernels for creating a filamentary density field.

The solution provided by the Bisous process is stochastic. It is thus expected that there is some variation in the detected filamentary patterns for different Markov-chain Monte Carlo (MCMC) runs of the model. However, thanks to the stochastic nature of the model, we gain a morphological and a statistical characterisation of the filamentary network simultaneously.

In practice, we first fix the width scale of the filaments to 1.4~Mpc since as found by \citet{2014MNRAS.438.3465T} this is the typical scale of the 
filaments in the cosmic web as traced by galaxies. The results are not very sensitive to the exact value of the scale. By applying the algorithm we then obtain a three dimensional ``visit'' map containing information related to the filament detection probability \citep[see][for the definition of this parameter]{2014MNRAS.438.3465T}. At the moment we do not have a method to relate this information to an exact detection probability. Rather, we set experimentally the lower limit for the visit parameter to 0.005 since this choice removes the noisy outer regions of the filaments and since the higher values would break filamentary structures clearly identified by eye inspection into disconnected sub-units (see Section~\ref{application}).

\subsection{Luminosity density field}
\label{ld_meth}

The goal of the luminosity density (LD) method is to build up a three dimensional luminosity density field using the spatial distribution and the luminosities of a given galaxy sample \citep[see][for details]{2012A&A...539A..80L}. In case of the observational data (in contrast to the simulated data) we start from the fluxes of galaxies which we convert into luminosities (including the $K$-corrections).

Since the 2dF galaxy catalogue is flux limited (to b$_{j}~\approx~19.5$) the mean galaxy number density and luminosity decrease with the redshift. 
To account for this selection effect we assign a statistical weight to each galaxy proportional to the fraction of the galaxy luminosity accessible to 
observations at a given redshift \citep[see][]{2011PhDT........68T, 2014A&A...566A...1T}.

We then smooth the galaxy luminosity distribution (observed or simulated) with a symmetric B3-spline kernel function. The smoothing scale sets the minimum physical scale of the structures that we can identify in our analysis. In this case we set it equal to 1.4~Mpc to match the characteristic scale of the filaments in the galaxy distribution \citep[see][]{2014MNRAS.438.3465T}.

We then sample the smoothed LD distribution at points of a uniform cubic grid with a grid size 1.4~Mpc\footnote{In order to avoid under-sampling, the sampling scale should not exceed our adopted smoothing scale.}, encompassing the survey volume, thus creating the LD field. A smaller sampling scale would be desirable, but computationally expensive in case of large simulations (see below). Our adopted sampling scale of 1.4~Mpc represents an acceptable compromise between the resolution and the computational cost. Note that when we analyse the individual sight-lines in the observational data we increase the sampling resolution in order to derive smooth distribution of the luminosity density (see Section~\ref{sculptor}).

\section{Large-scale structure simulation}
\label{simu}
As discussed in Introduction, numerical simulations of the build-up of the cosmic structures, as well as theoretical arguments, indicate that both galaxies and the WHIM trace the same underlying distribution of dark matter. 
The relation between the baryon tracers and the underlying mass is not deterministic since it also depends on the type of tracer, the scale, the 
underlying density and the  ill-known stellar feedback processes. It also depends on time. In our analysis we consider a rather local sample, so the latter
dependency can be ignored.

However, in the filamentary regions, i.e. on scales significantly larger than those affected by stellar processes, the LD--WHIM density relation may be usefully tight. Building upon this assumption, we shall consider the hydrodynamical simulations carried out by \citetalias{2012MNRAS.423.2279C} to derive a relation between the galaxy LD and the WHIM density within cosmic filaments. The procedure to build up and calibrate this relation is described in this Section.

\subsection{Gas and dark matter}
  
The simulations of \citetalias{2012MNRAS.423.2279C} have been carried out using the TreePM-smoothed particle hydrodynamics (SPH) code GADGET-3, an improved version of GADGET-2 \citep{2005MNRAS.364.1105S}. It follows the evolution of a box with a co-moving size of 590~Mpc per side considering both dark matter and gas. Each component is described using $1024^3$ particles from $z=41$ to the present epoch (see \citetalias{2012MNRAS.423.2279C} for full details).

We used these particle data as an input to our analysis. We divided the full simulation volume into a cubic grid whose grid size was determined from two criteria: the minimum gas overdensity that we want to sample and the need to avoid oversampling the gas density field. Since we focus on the WHIM, we want to trace this gas down to a density contrast $\delta_\mathrm{b} = 1$. This corresponds to an inter-particle separation of 0.6~Mpc in the simulation, which is thus the minimal useful grid size. In order to avoid the oversampling of the field we set the grid size to 1.4~Mpc, for both the gas and the DM. With this choice, the gas and DM density field of the simulation are defined on a $410^3$ cubic grid which constitutes the input data set to model the gas properties.

The physical modelling of the gas component includes radiative cooling, star formation and feedback from supernova remnants, but not AGN feedback. Even if ignoring AGN feedback is rather unphysical, this assumption does not affect our work since the galaxy luminosity is self-consistently modelled in the framework of the halo occupation distribution model (see next Section) whereby galaxy properties are determined from the DM distribution on a statistical basis. In addition, since we shall focus on a rather local galaxy sample within which evolutionary effects can safely be ignored, we only consider the $z=0$ output of the simulation.

We then created 3D maps of the gas density and temperature and the dark matter density defined on a cubic grid as follows. Each gas particle's mass, $m_i$, is distributed over a variable number of cells with weight proportional to the SPH smoothing kernel integrated over the cell volume (see similar applications in \citet{2006MNRAS.368...74R, 2007MNRAS.378.1259R, 2012MNRAS.424.1012R}. The gas density is then computed by dividing the mass associated to each cell by its volume. We then computed the mass-weighted gas temperature 3D-maps by smoothing the value of $m_i T_i$ of each gas particle, where $T_i$ is the SPH particle's temperature, and then dividing each cell's value by its mass obtained from the previous mapping.

DM particles are treated differently. The DM density field is interpolated at the points of the same cubic grid as above, using the Clouds-In-Cells (CIC) method that assigns each DM particle to 8 cells with weight that depends on the particle position.

\subsection{Galaxies}
\label{galaxies}
We created the galaxies for the above simulation \citepalias{2012MNRAS.423.2279C} 
by populating the DM halos using a halo occupation distribution (HOD) constrained with the SDSS data \citep{2011ApJ...736...59Z}. In summary, we used the masses and positions of each DM halo in order to obtain the magnitudes and positions of the galaxies as follows (see Appendix~\ref{app:a} for more details):
1) in order to know the spatial scale of each halo, we defined its virial radius according to the spherical collapse formalism \citep[see e.g.][]{1980lssu.book.....P, 1996MNRAS.282..263E, 1996MNRAS.280..638K, 1998ApJ...495...80B};
2) we populated each halo with subhalos, up to the virial radius, by performing a Monte Carlo realisation of the subhalo mass function model \citep{2010MNRAS.404..502G};
3) for each halo we performed an abundance matching approach to assign a given central galaxy or a satellite with a luminosity using halo occupation distribution of \citet{2011ApJ...736...59Z}.

Our HOD implementation predicts galaxy luminosities in the $r$ band, whereas we wish to apply the results to 2dF galaxies (see below) that were observed in the $b_j$ band. However, one can extrapolate LD predictions from a band $x$ to another band $y$, if the the mean LD has been estimated in both bands: assuming that the luminosity overdensity $\delta_\mathrm{LD}=\mathrm{LD}/\langle \mathrm{LD}\rangle$ for a given galaxy population is band-independent, one can estimate $\mathrm{LD}_{y}=\mathrm{LD}_{x}\times\langle\mathrm{LD}_{y}\rangle/\langle\mathrm{LD}_{x}\rangle$. The mean LD value at $z=0$ in the $r$-band obtained from the \citetalias{2012MNRAS.423.2279C} simulation and that in the $b_j$ band obtained from 2dF (both estimated by us) are listed in Table~\ref{conv.tab}.

In addition, one should also consider the effect of morphological segregation, i.e. the fact that red galaxies populate mostly the high density regions and blue galaxies are more homogeneously distributed. Since we are interested in the WHIM filaments, i.e. we concentrate on the low to intermediate overdensity regions, our galaxy populations are expected to be dominated by blue galaxies. Indeed, based on our preliminary analysis, the morphological segregation of galaxies in filaments is very weak.

\begin{table}
 \centering
  \caption{Conversion factors.
 \label{conv.tab}}
  \begin{tabular}{lc}
  \hline\hline
name                                         & value  \\   
\hline\hline
$\langle\mathrm{LD}_{b_j}\rangle$\tablefootmark{a} &   $0.0113\times 10^{10}L_{\odot}$~Mpc$^{-3}$ \\
$\langle\mathrm{LD}_{r}\rangle$\tablefootmark{b}     &  $0.00772\times 10^{10}L_{\odot}$~Mpc$^{-3}$ \\
$\langle\rho_\mathrm{b}\rangle$\tablefootmark{c}     &   $0.618\times 10^{10}M_{\odot}$~Mpc$^{-3}$  \\
redshift CMB correction\tablefootmark{d}      & $-9.4\times 10^{-4}$ \\
\hline
\end{tabular}
\tablefoot{
\tablefoottext{a}{The mean galaxy luminosity density of the full 2dF area in the $b_j$ band. }
\tablefoottext{b}{The mean galaxy luminosity density of the \citetalias{2012MNRAS.423.2279C} simulation in the $r$ band. }
\tablefoottext{c}{The cosmic baryon density at $z = 0$. }
\tablefoottext{d}{To be added to heliocentric redshifts for conversion to CMB rest frame in direction of H2356-309.}
}
\end{table}

\begin{figure*}
	\centering
\vspace{-3cm}
\hbox{
\hspace{-0.8cm}
\includegraphics[width=10.5cm,angle=0]{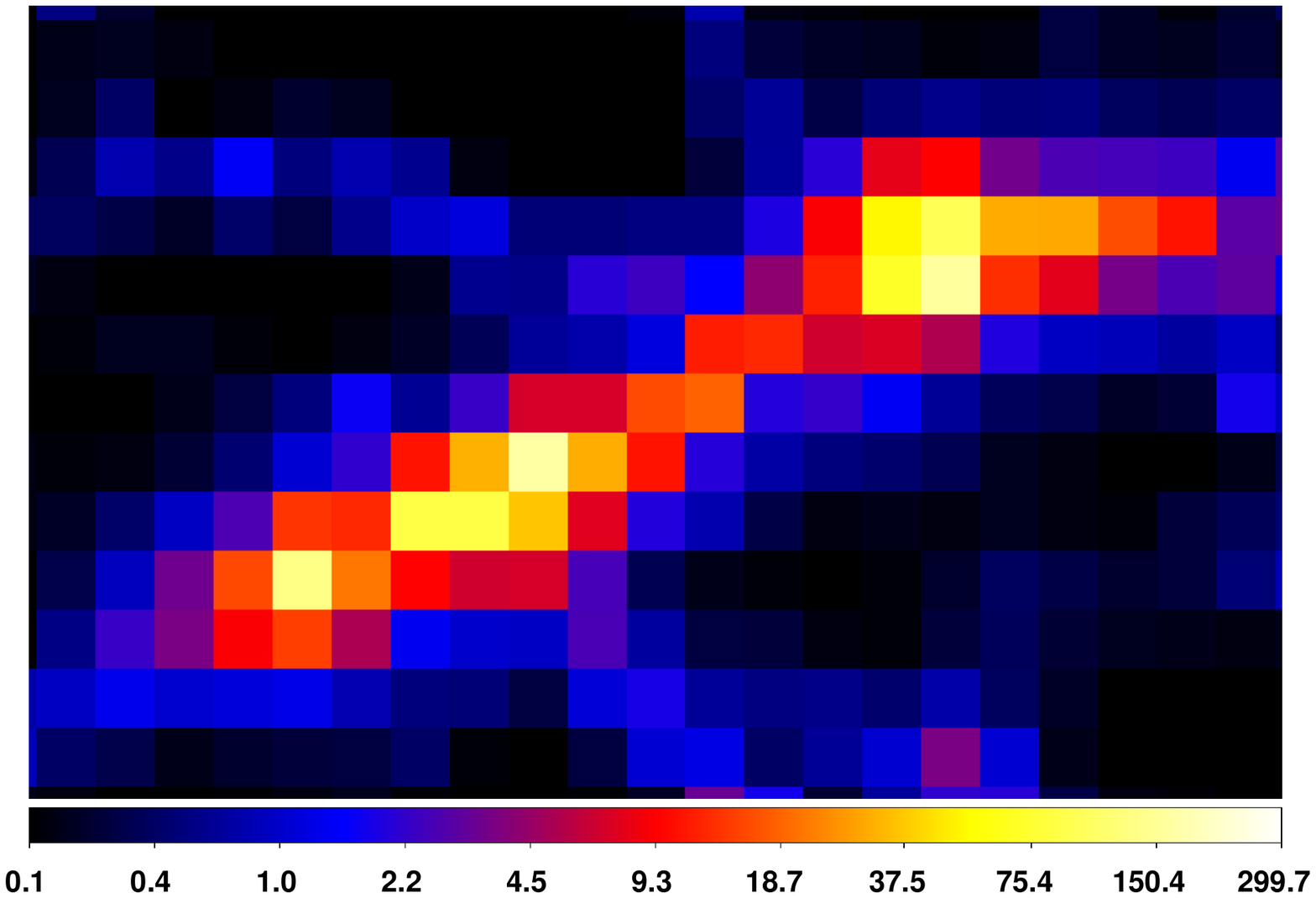}
\hspace{-1.5cm}
\includegraphics[width=10.35cm,angle=0]{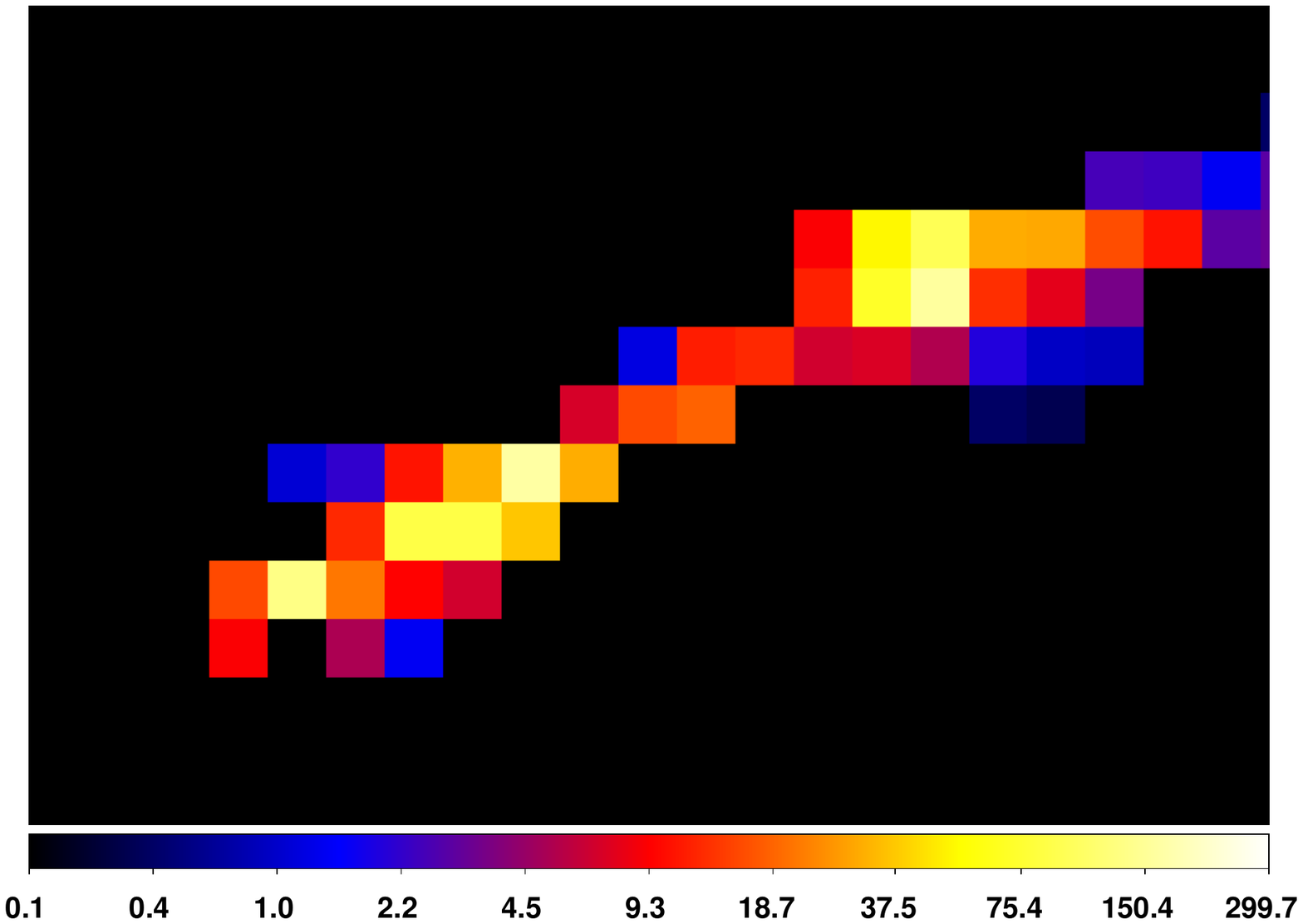}
}
\vspace{-3cm}
\caption{Gas density ($10^{10}~M_{\odot}\,\mathrm{Mpc}^{-3}$) image of an interesting region from the simulation \citepalias{2012MNRAS.423.2279C} when applying no filtering (\emph{left panel}) and when applying the filamentary environment selection (\emph{right panel}). The pixel size and the image thickness are both 1.4~Mpc.}
\label{blob.fig}
\end{figure*}

\subsection{Application of filament-finding and LD field methods to simulated data}
\label{application}

We constructed the $r$ band luminosity density (LD$_r$) fields by applying the LD method to the galaxy population created above. We first extracted the luminosity densities and gas densities in the full volume of the simulation studied in this work. We found that the fraction of the gas mass in the WHIM temperature and density range, i.e. the WHIM volume filling factor, is $\sim$50\%.
 
We then applied the Bisous model to the galaxy distribution and defined volume elements with visit parameter values higher than 0.005 (see Section~\ref{bisous}) as members of filaments. The benefit of this procedure is that it can be applied both to simulated and observational data, based only on the 3D distribution of the galaxies. However, in the case of the observational data, we do not have any temperature or density information of the studied intergalactic gas. Thus, for consistency, we will not directly apply any temperature or density selection when extracting the simulated data.

The environmental selection did in fact have the desired effect of excluding the voids and the noisy, low temperature filament edges and concentrating more on the filament cores (see Fig.~\ref{blob.fig}). The WHIM mass fraction increased from 50\% to 70\% when we restricted the analysis to such environmentally-selected volume. In the following we thus assume that 70\% of the gas mass in the regions satisfying the Bisous model filament detection criteria is WHIM, i.e. $\rho_\mathrm{whim}$~=~$0.7~\times~\rho_\mathrm{gas}$.

We want to exclude galaxy clusters from the analysis since their high galaxy number density and low expected WHIM mass fraction \citep[e.g.][]{2009ApJ...697..328B} yields a very different LD--WHIM density ratio from that in the filaments. The above filtering minimises rather efficiently the contribution of clusters of galaxies, proven by the fact that only $\sim$10\% of the selected gas mass has temperature and densities that both exceed the WHIM upper limits.

\begin{figure*}
	\centering
\vspace{-1.5cm}
\vbox{
\hbox{
\includegraphics[width=9.0cm,angle=0]{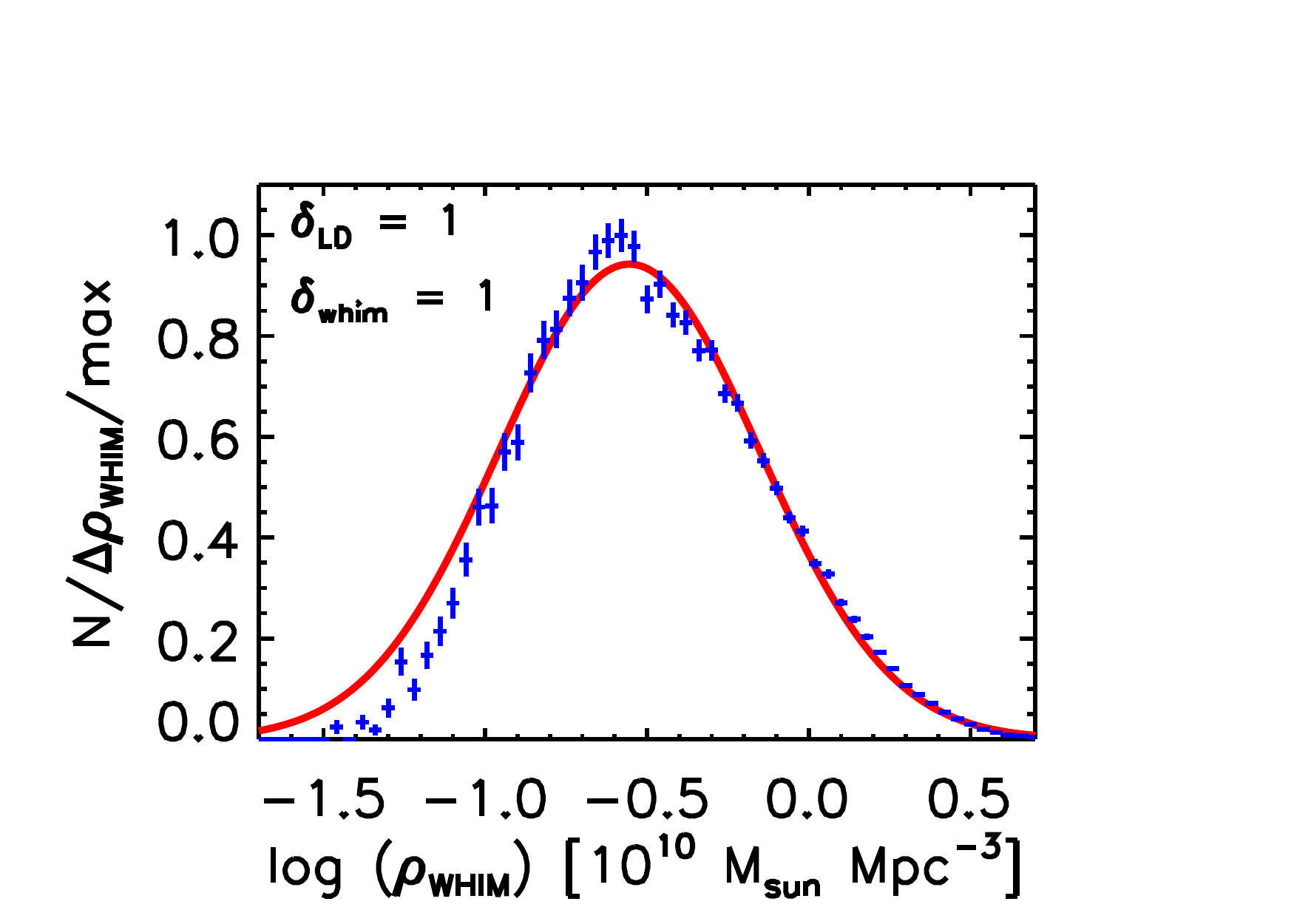}
\includegraphics[width=9.0cm,angle=0]{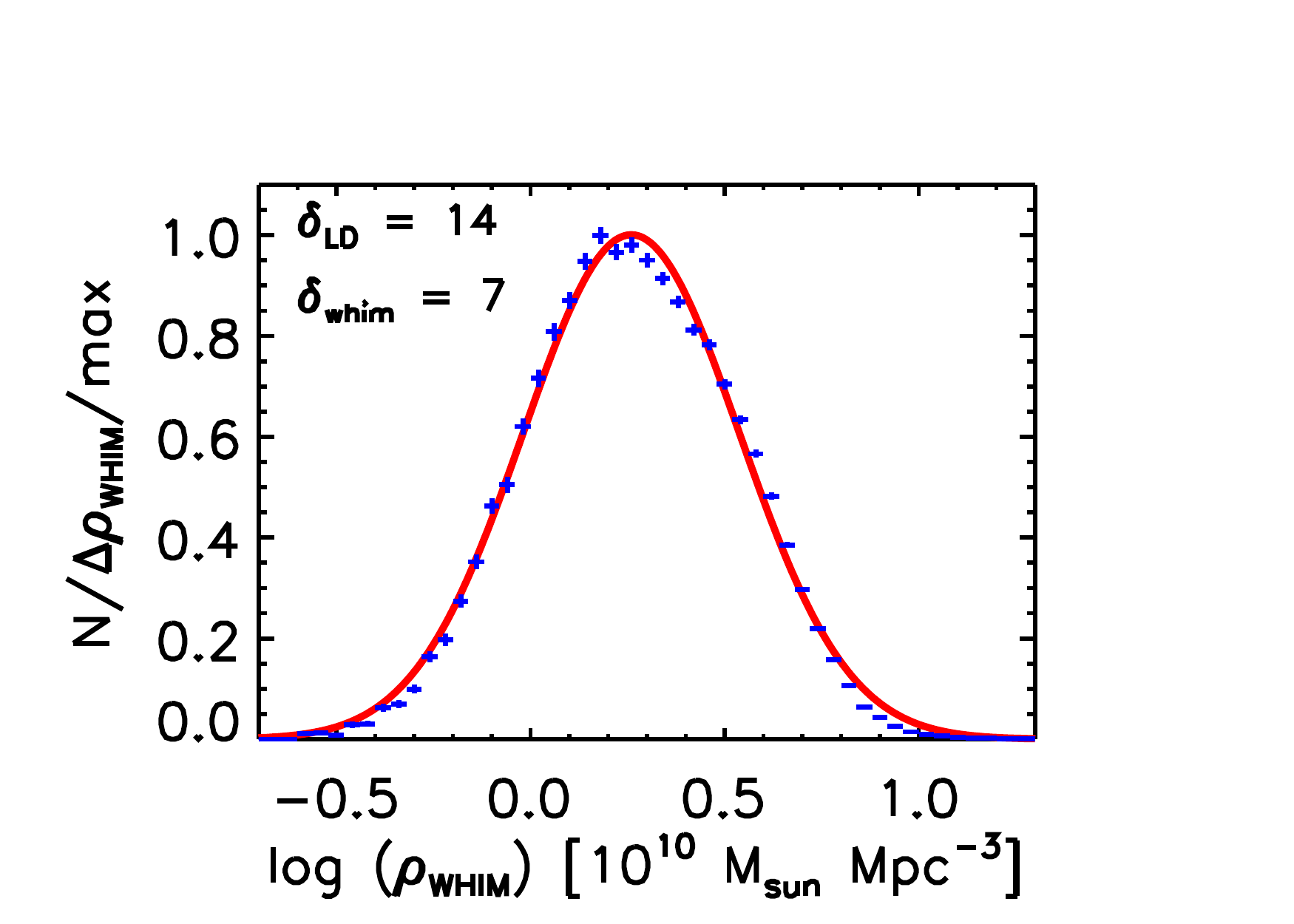}
\vspace{-1.0cm}
}
\hbox{
\includegraphics[width=9.0cm,angle=0]{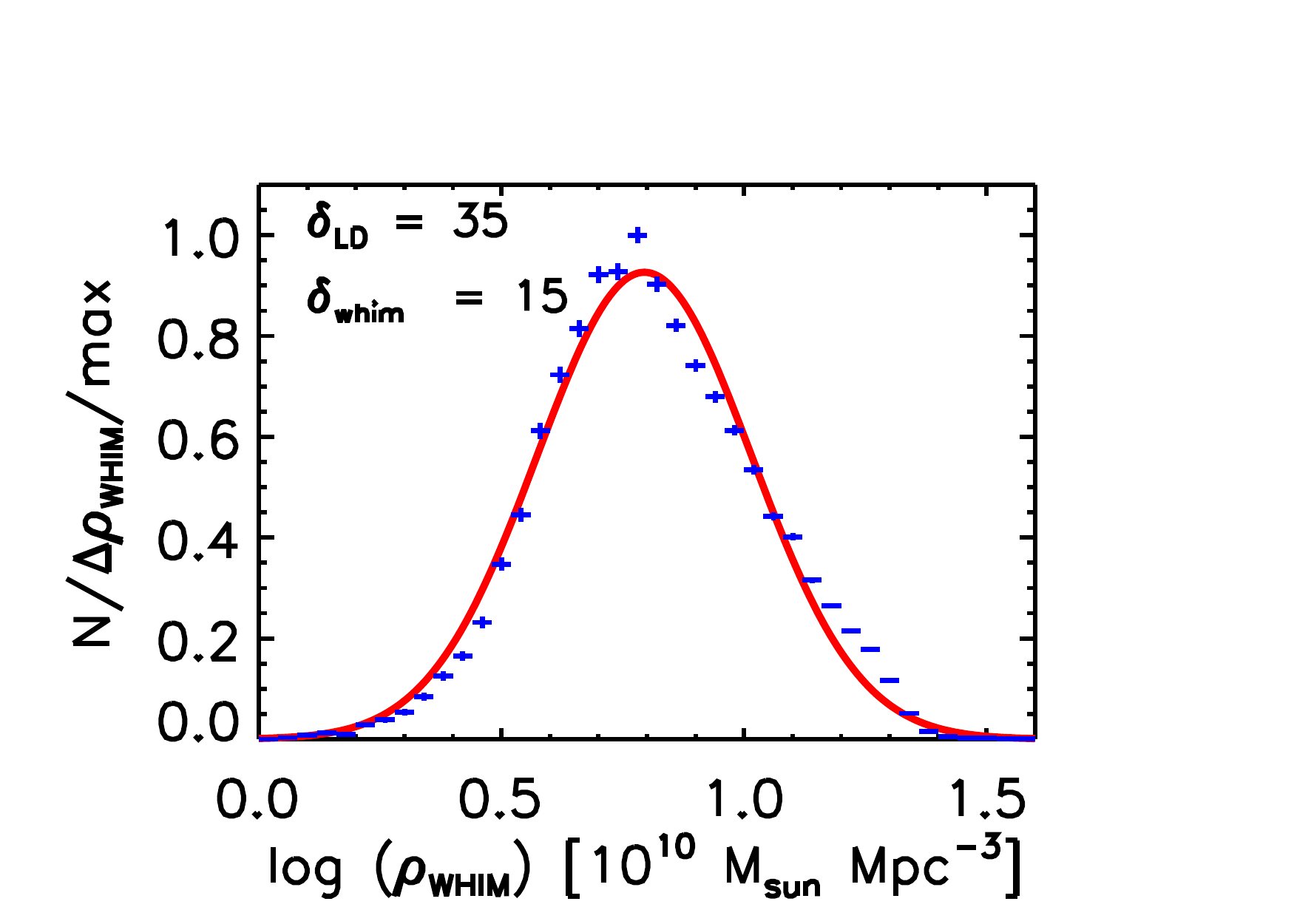}
\includegraphics[width=9.0cm,angle=0]{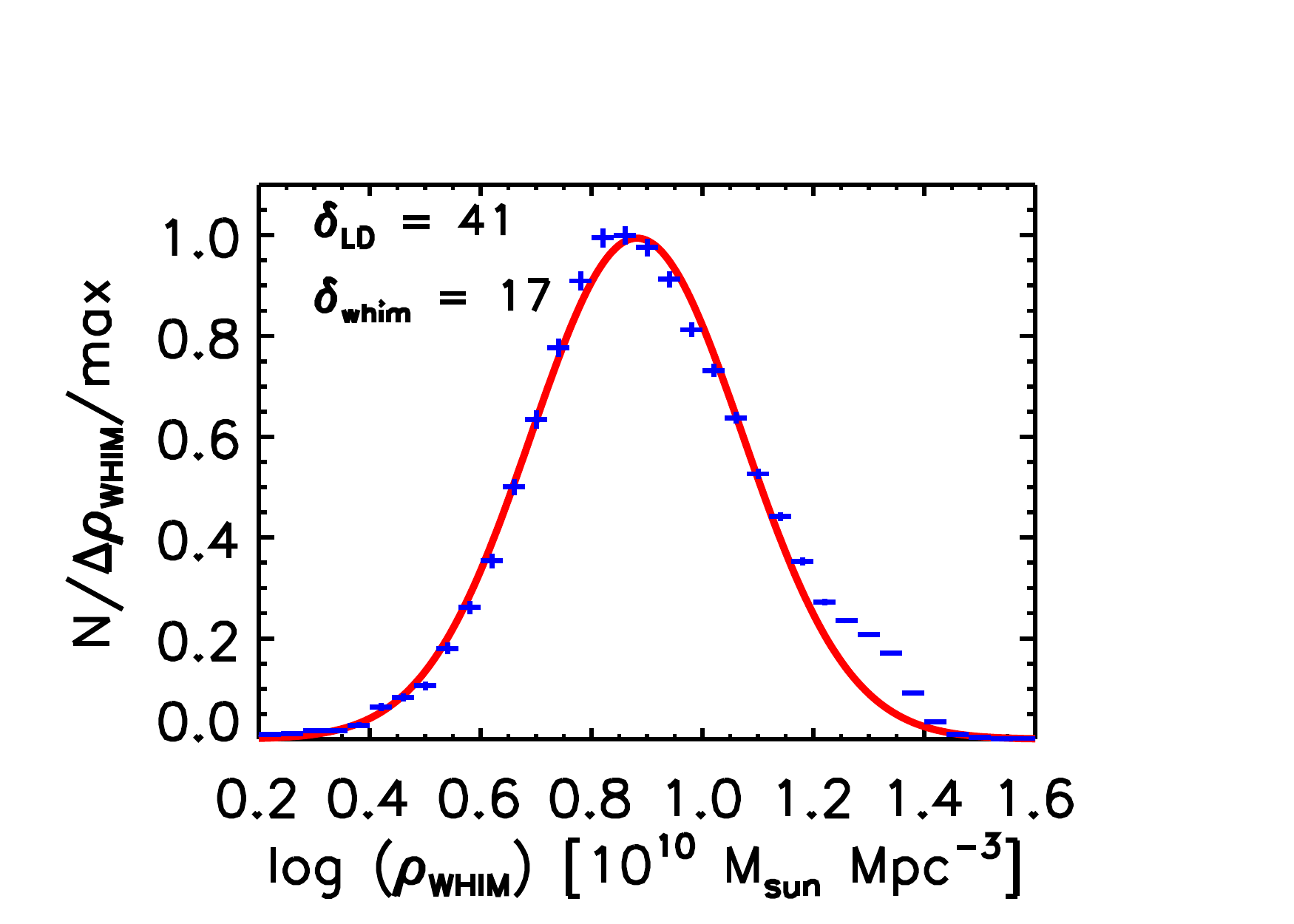}
\vspace{-1.0cm}
}
\hbox{
\includegraphics[width=9.0cm,angle=0]{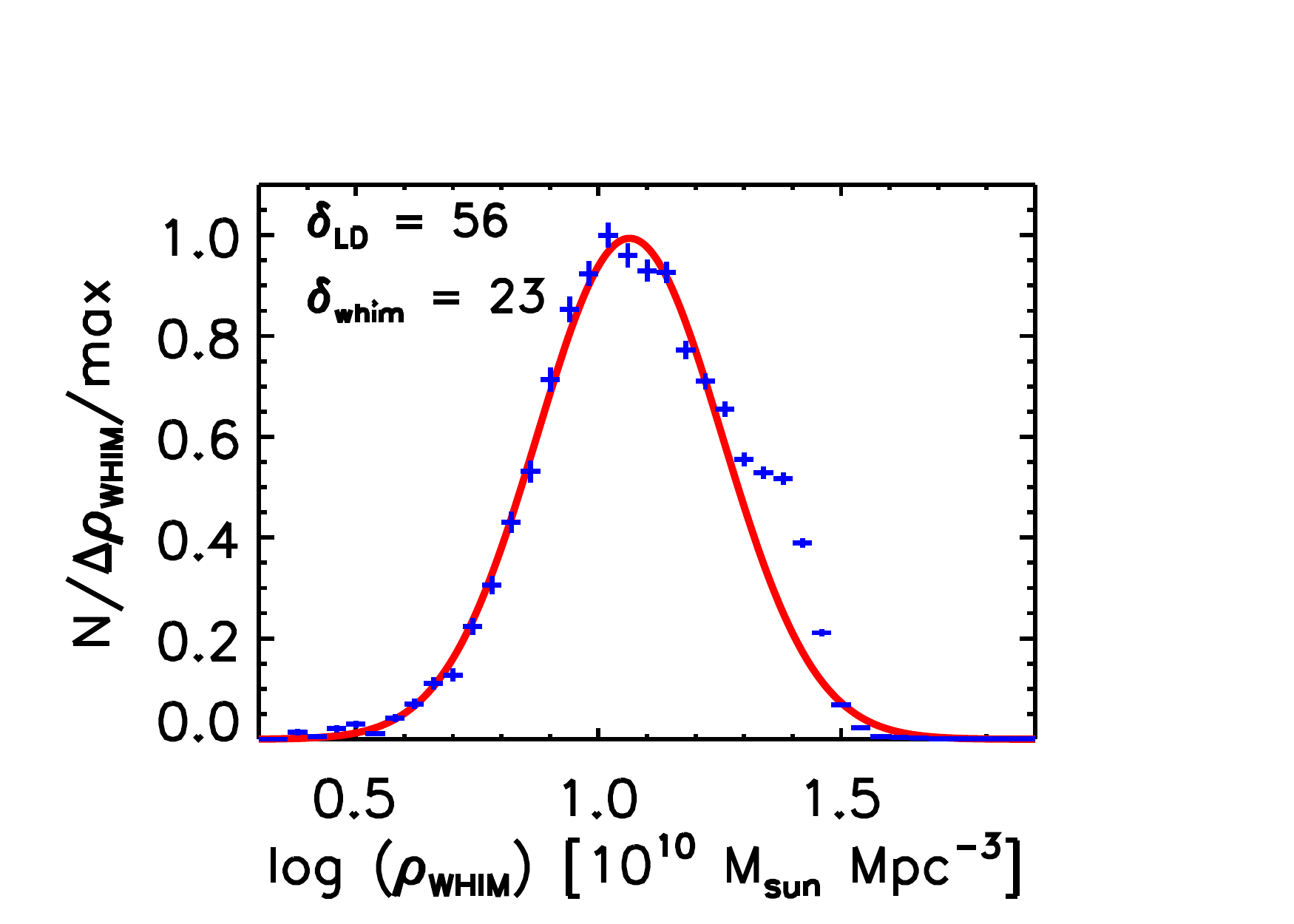}
\includegraphics[width=9.0cm,angle=0]{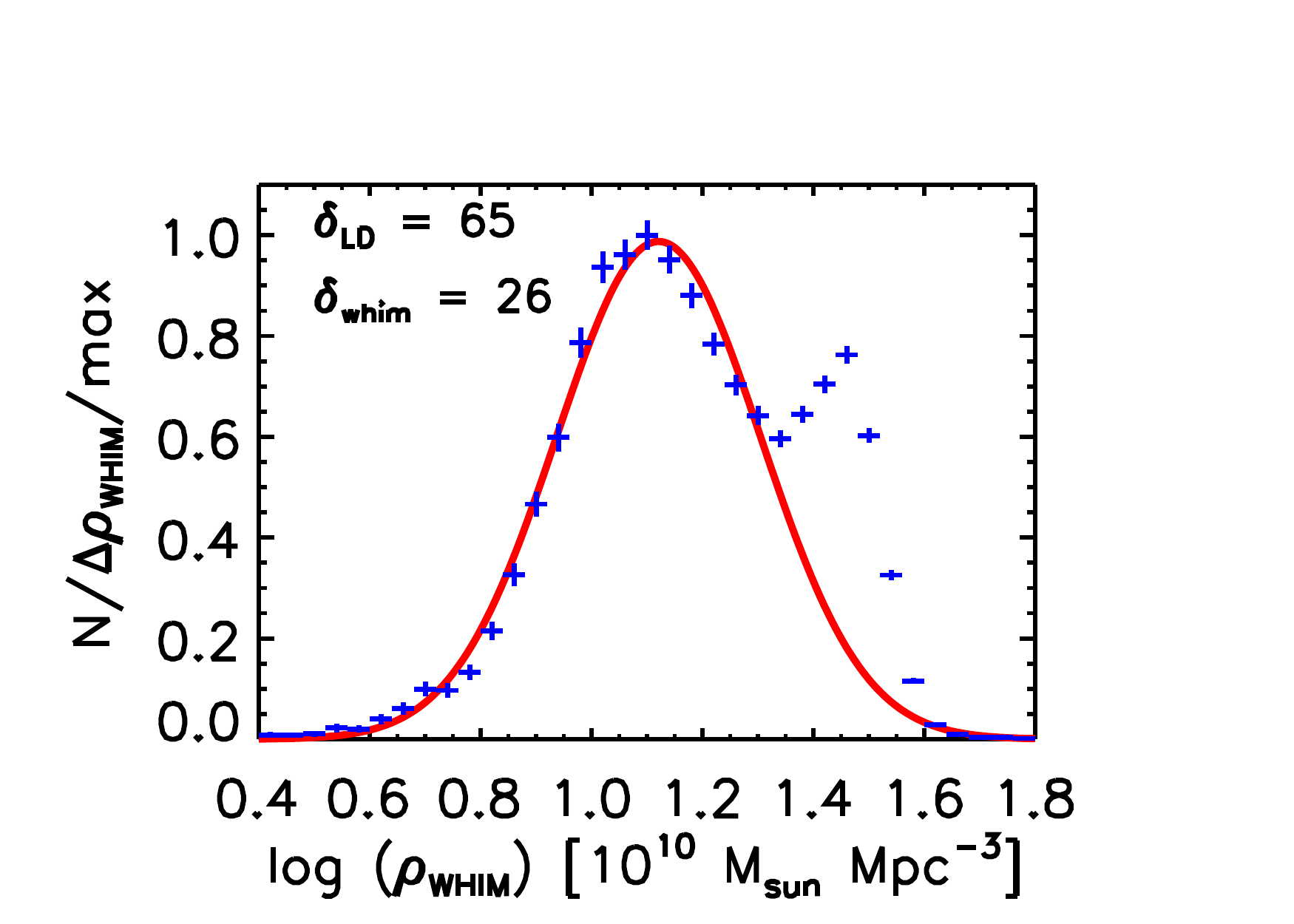}
\vspace{-1.0cm}
}
\hbox{
\includegraphics[width=9.0cm,angle=0]{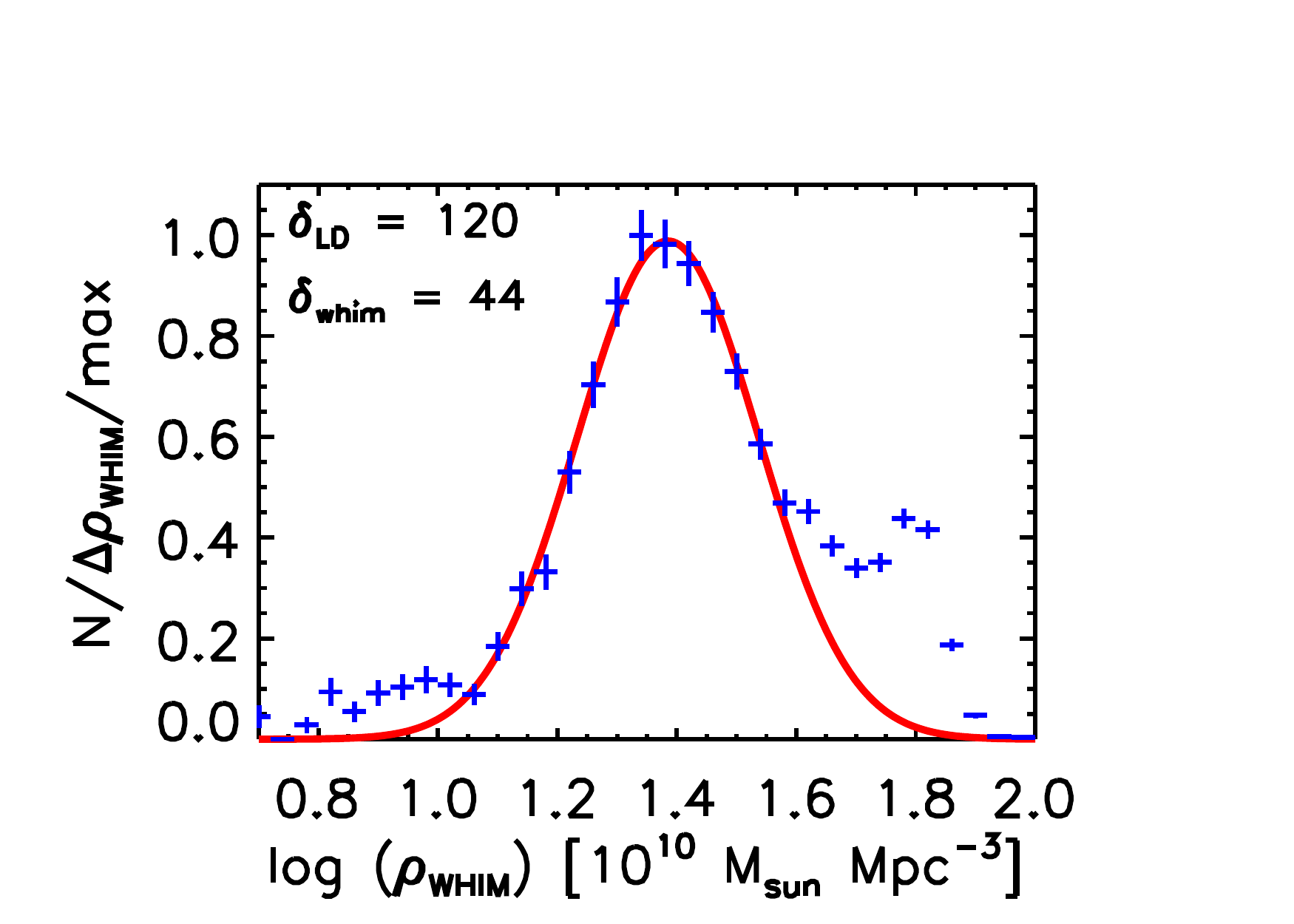}
\includegraphics[width=9.0cm,angle=0]{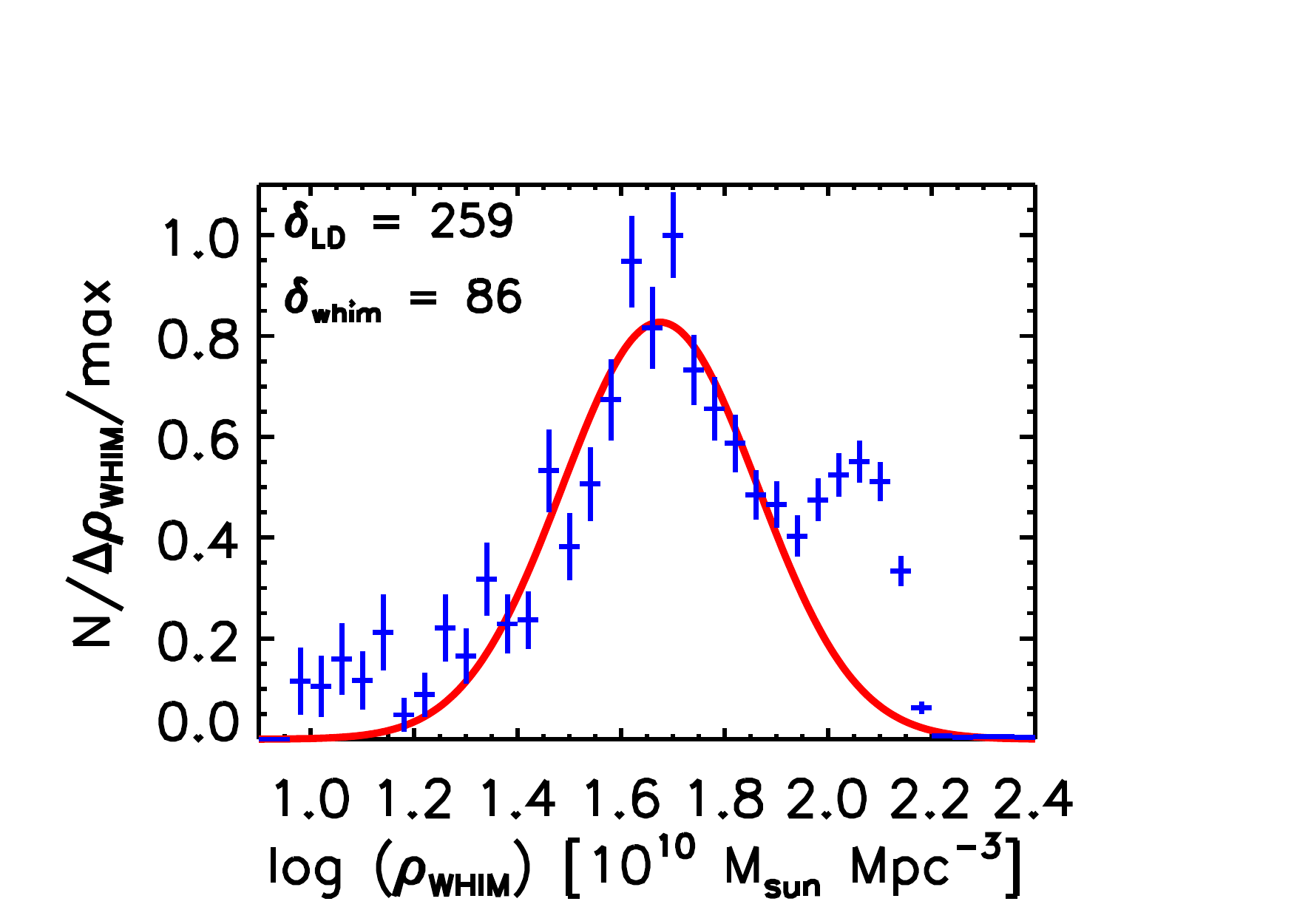}
}
}
\caption{The WHIM density distributions (crosses) and the best-fit log-normal models (lines) within selected LD bins. $\delta_\mathrm{LD}$ indicates the central luminosity overdensity value of a given bin.  $\delta_{\rm whim}$ indicates the WHIM overdensity value corresponding to the peak value.}
\label{WHIMdistr.fig}
\end{figure*}

\subsection{Determining the PDF of the WHIM density in different luminosity environments}
\label{WHIM_ld_distr}

The relation between the WHIM density and the luminosity density, both stochastic tracers of the underlying mass density field is, unsurprisingly, also  stochastic. As a result, for a given LD$_r$ the values of the WHIM density spread over quite a broad range. Here we wanted to quantify more than the simple spread: we characterised the constrained probability function $P(\rho_{\rm whim}~|~\mathrm{LD}_r)$. 

We have restricted our analysis in the LD$_r$ range of [0.01--2.0]~$\times~10^{10}L_{\odot}$\,Mpc$^{-3}$. This choice is justified on the basis of the 
simulations of \citetalias{2012MNRAS.423.2279C} that indicated that $\sim$~95\% of the WHIM mass of the full simulation volume is within this LD$_r$ 
range. Increasing the range would therefore decrease the “purity” of the sample.
We divided this LD$_r$ range into 15 equally-spaced logarithmic bins and 
measured the conditional PDF in each of them.

The frequency histograms of the WHIM gas density at a given LD$_r$ bin constitute our estimate of the conditional probability distribution function $P(\rho_{\rm whim}~|~\mathrm{LD}_r$) (see Fig.~\ref{WHIMdistr.fig}). The WHIM density value corresponding to the position of the maximum correlates with the LD$_r$ value, which indicates the presence of a LD--WHIM density correlation that we explore in the next Section. As long as LD$_r$ is small the PDF is well approximated by a log-normal distribution. For $\delta_\mathrm{LD}>10$ a positive skewness appears that develops into a secondary peak at $\delta_{\rm whim}~\sim 30$ which shifts to progressively larger $\rho_{\rm whim}$ values when LD increases.
 These features suggest the presence of a population of luminous objects with associated WHIM gas with over-densities higher than $\sim 10$. We speculate that these objects correspond to luminous galaxies or groups with relatively dense WHIM gas in their X-ray halos. The fact that the WHIM gas associated to this peak seems to follow the same LD--WHIM density relation as the gas in the main peak indicates that this is a real feature and not a simulation artifact.

\begin{figure*}
	\centering
\vspace{-1.0cm}
\hbox{
\includegraphics[width=10.0cm,angle=0]{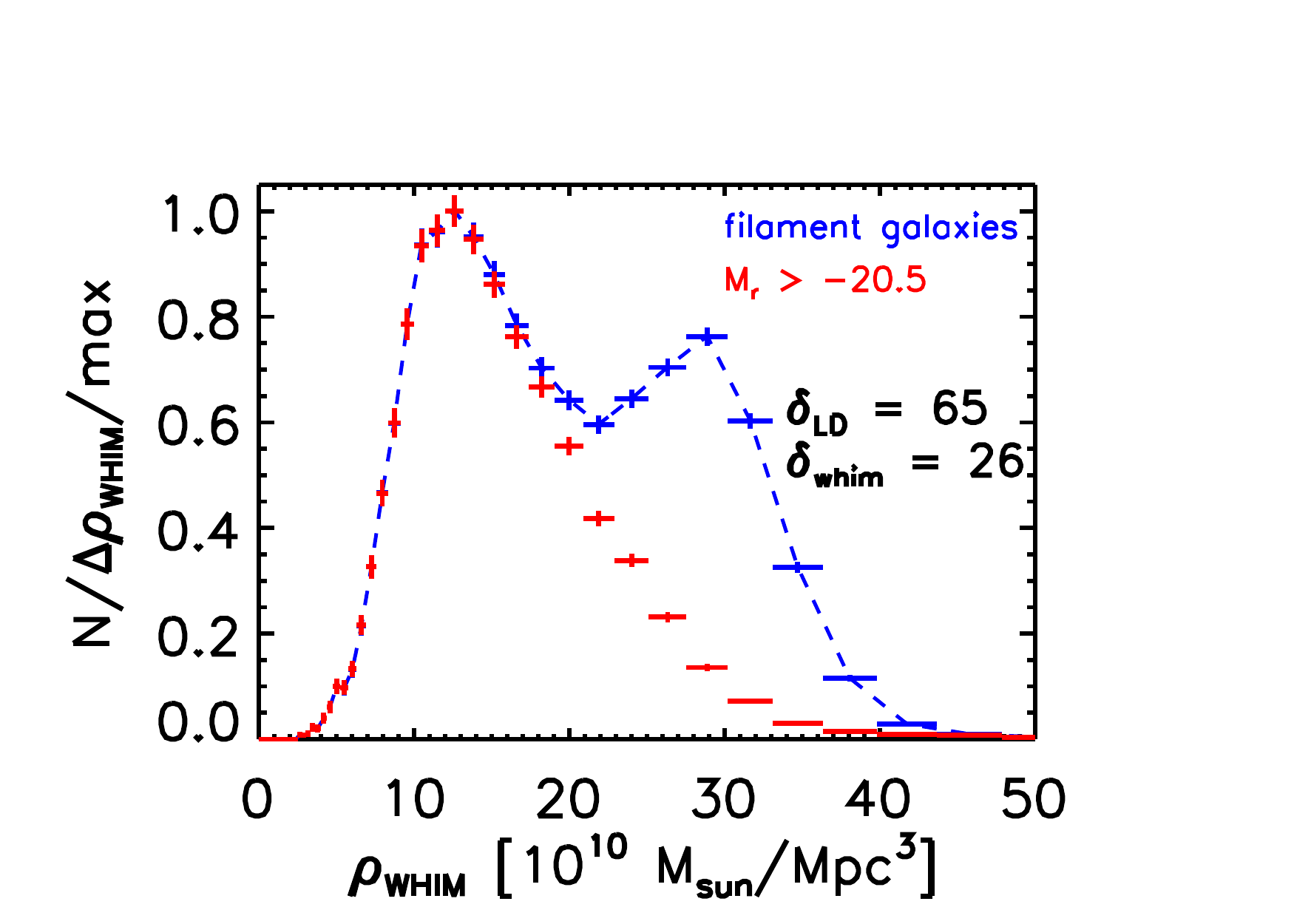}
\hspace{-2.5cm}
\includegraphics[width=10.0cm,angle=0]{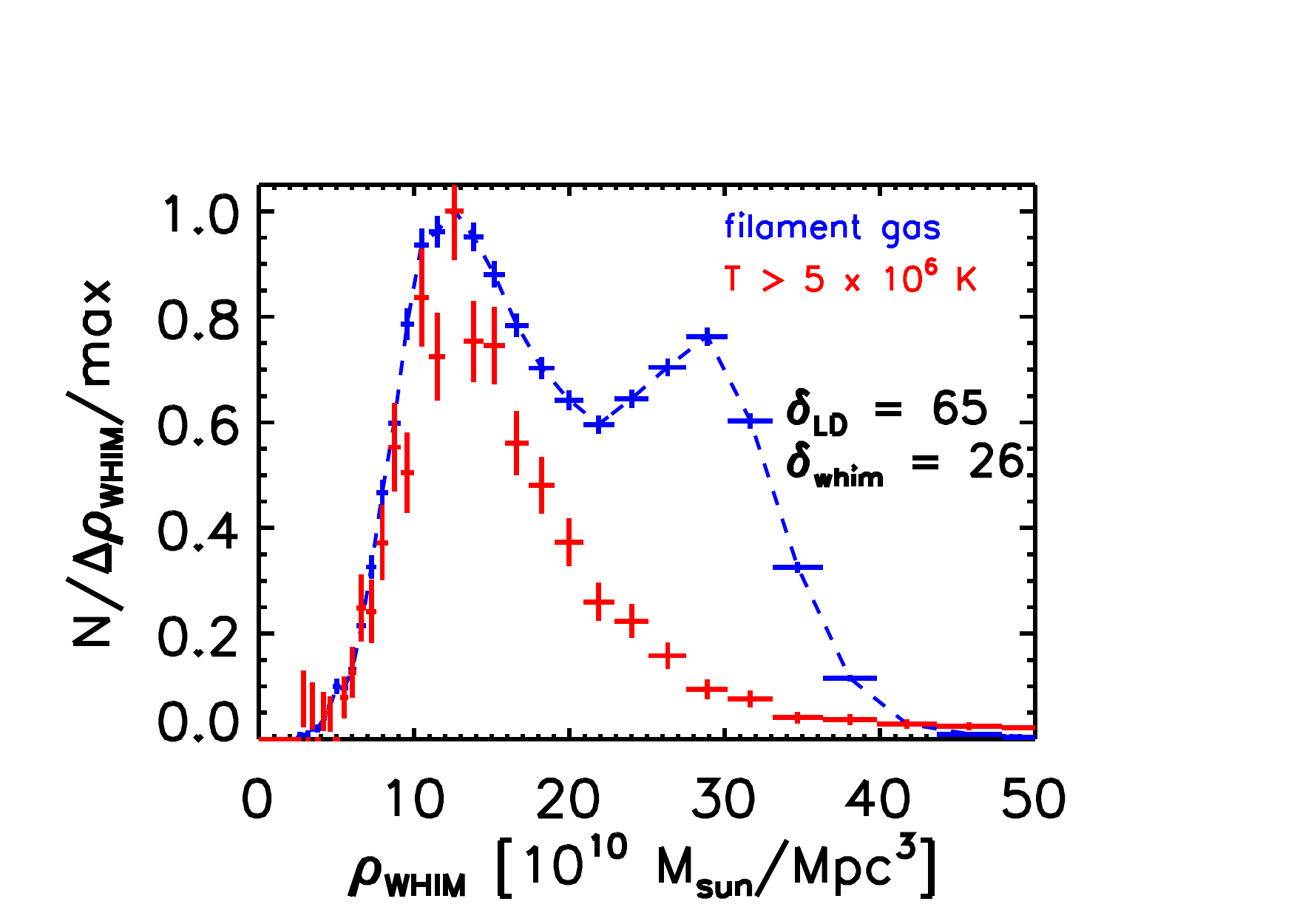}
}
\caption{The WHIM density distribution at $\delta_\mathrm{LD}~\sim~65$ (where the WHIM density peaks at $\delta_\mathrm{b}~\sim$26) using all the simulation data within the filamentary environments are shown with blue crosses and dashed lines. Red crosses indicate the distribution when using only volume 
elements with galaxies fainter than $M_r~=~-20.5$ (left panel) or with gas hotter than T~=~$5~\times~10^6$~K}.
\label{secondary.fig}
\end{figure*}

To test this hypothesis we went back to the simulations to characterise the mock galaxies and the gas that contribute to the secondary peaks. We found that the peak disappeared when we removed from the sample the data from volume elements with galaxies brighter than $M_r = -20.5$ or gas colder than T~=~5$\times~10^6$~K (see Fig.~\ref{secondary.fig}). This indicates that the second peak is contributed by gas particles preferentially associated to bright galaxies in halos with masses in excess of a few times $10^{12}~M_{\odot}$ (see Appendix~\ref{app:a}). These objects are rarely isolated and instead are typically found in groups (see Appendix~\ref{app:a}). 
Thus, and considering also the relatively low temperature, the secondary peak appears to be contributed by gas halos of galaxies within galaxy groups or by the intra-group gas, or both.

\subsection{LD -- WHIM density relation}
\label{rel.sec}

Within the filamentary environments, the simulated WHIM density and the LD$_r$ correlate well: the Pearson correlation coefficient is $\sim$0.80 while the number of data points is $\sim~1.7~\times~10^6$. This indicates that the LD traces well the WHIM. Thus, we utilise in the following this correlation for finding the WHIM.

We obtained our main goal of estimating the relation between LD and WHIM gas density as follows. 
1) We considered the conditional probability functions  $P(\rho_{\rm whim}~|~\mathrm{LD}_r$) in each one of the 15~LD$_r$ bins.
2) We Monte Carlo sampled the individual PDFs to obtain a statistical realisation of the LD$_r - \rho_{\rm whim}$ relation.
3) We fitted the corresponding data with a power-law relation $\rho_{\rm whim} = A\times \mathrm{LD}_r^{B}$ and stored the best fit parameters $A$ and $B$.
4) We went back to step~1 and repeated the procedure $10^4$ times.
5) We estimated the resulting PDFs for $A$ and $B$.

\begin{figure*}
	\centering
\hbox{
\hspace{-0.5cm}
\includegraphics[width=10cm,angle=0]{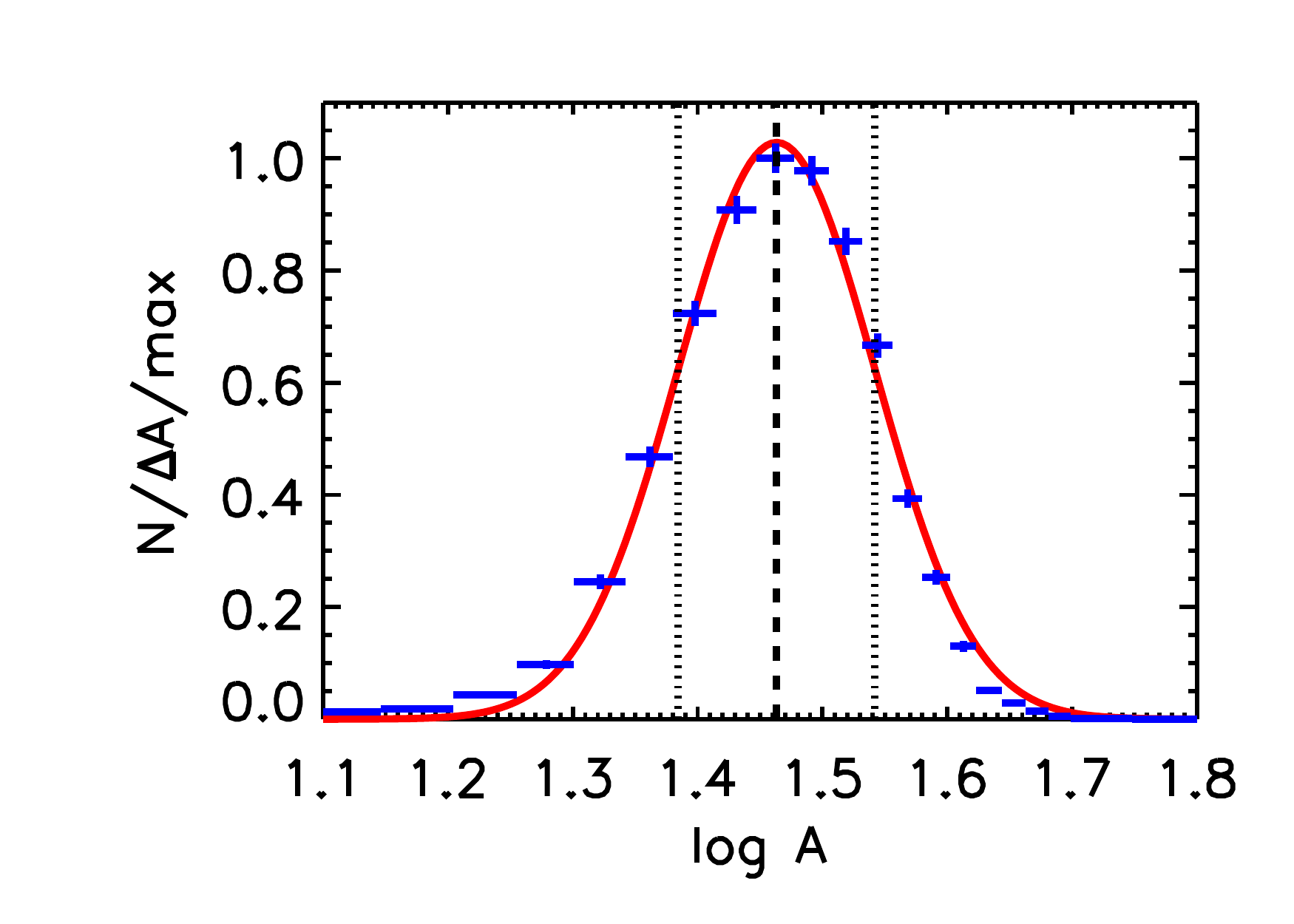}
\hspace{-1.5cm}
\includegraphics[width=10cm,angle=0]{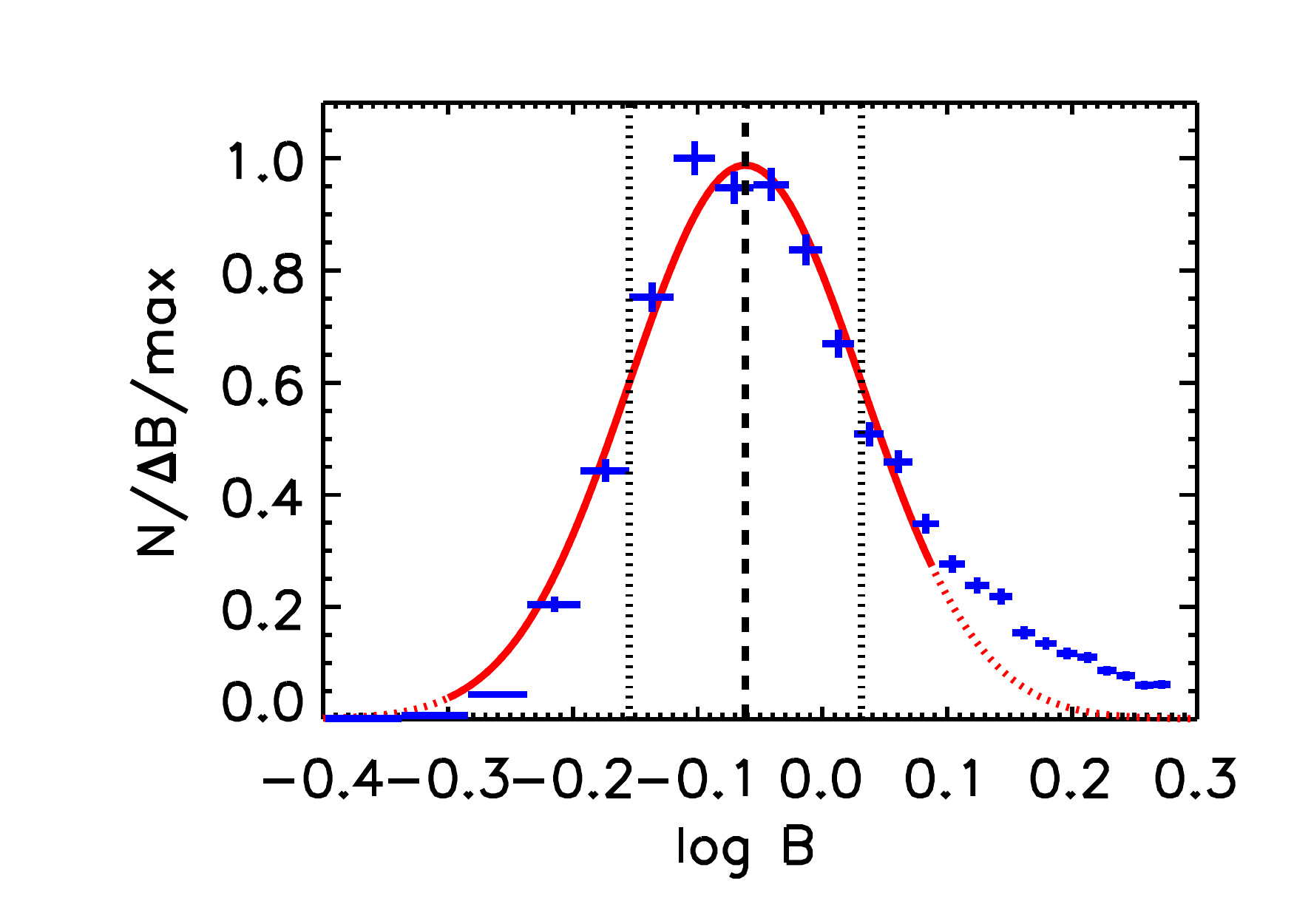}
}
\vspace{-0.5cm}
\caption{The distributions (blue crosses) of the power-law $\rho_{\rm whim} = A\times \mathrm{LD}_{r}^{B}$ normalisation $A$ (left panel) and the index $B$ (right panel) when fitting the randomised LD$_r$-WHIM density relations. The best-fit log-normal models are shown with red solid line. The dotted red line shows the extrapolation of the best-fit model. The centroid and 68\% intervals are indicated with the dashed and dotted vertical lines.}
\label{norm_index.fig}
\end{figure*}

For both parameters the PDF was well approximated by a log-normal distribution except in the positive tail in which the $B$ parameters exhibit an excess probability 
(see Fig.~\ref{norm_index.fig}). Using the centroids and the widths of the best-fit log-normal models to the distributions we thus obtained the 
LD$_r - \rho_{\rm whim}$ relation as
\begin{equation}
\rho_{\rm whim}   = 0.4 \pm 0.1  \times 10^{10}~M_{\odot}~\mathrm{Mpc}^{-3} \times {\delta_\mathrm{LD}}^{0.9 \pm 0.2} 
\label{rel1.eq}
\end{equation}
and an  analogous one expressed in terms of gas overdensity
\begin{equation}
\delta_{\rm whim} = 0.7 \pm 0.1  \times {\delta_\mathrm{LD}}^{0.9 \pm 0.2}, 
\label{rel2.eq}
\end{equation}
where the luminosity overdensity  $\delta_\mathrm{LD}~=~\mathrm{LD}/\langle \mathrm{LD} \rangle$ is the independent variable,
$\delta_{\rm whim}~=~\rho_{\rm whim}/\langle\rho_\mathrm{b}\rangle$, and  $\langle\rho_\mathrm{b}\rangle~=~0.62~\times~10^{10}~M_{\odot}~{\rm Mpc}^{-3}$ at $z=0$ \citep{2015arXiv150201589P}.

\begin{figure*}
	\centering
\hspace{-1.0cm}
\hbox{
\includegraphics[width=10cm]{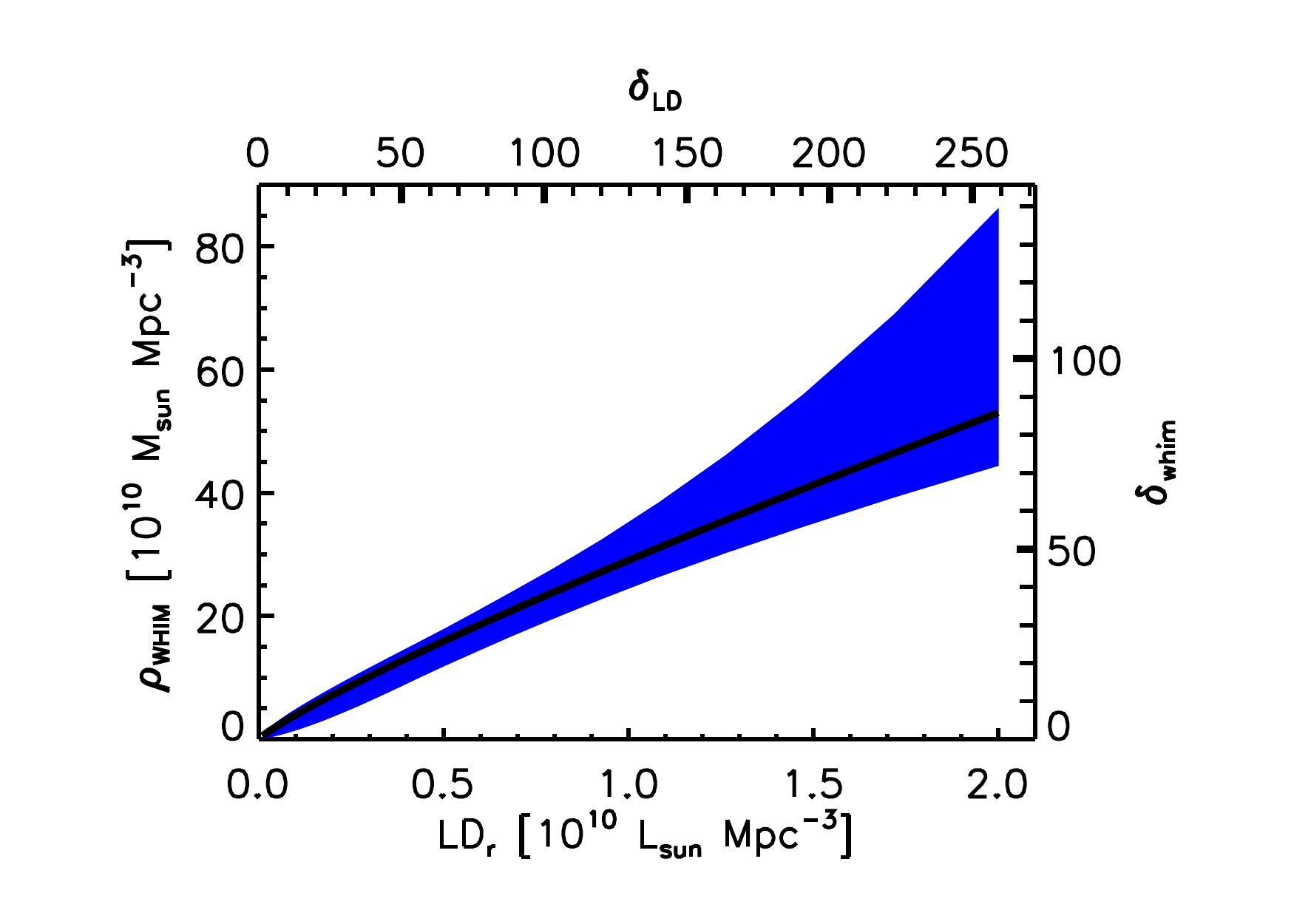} 
\hspace{-2.0cm}
\includegraphics[width=10cm]{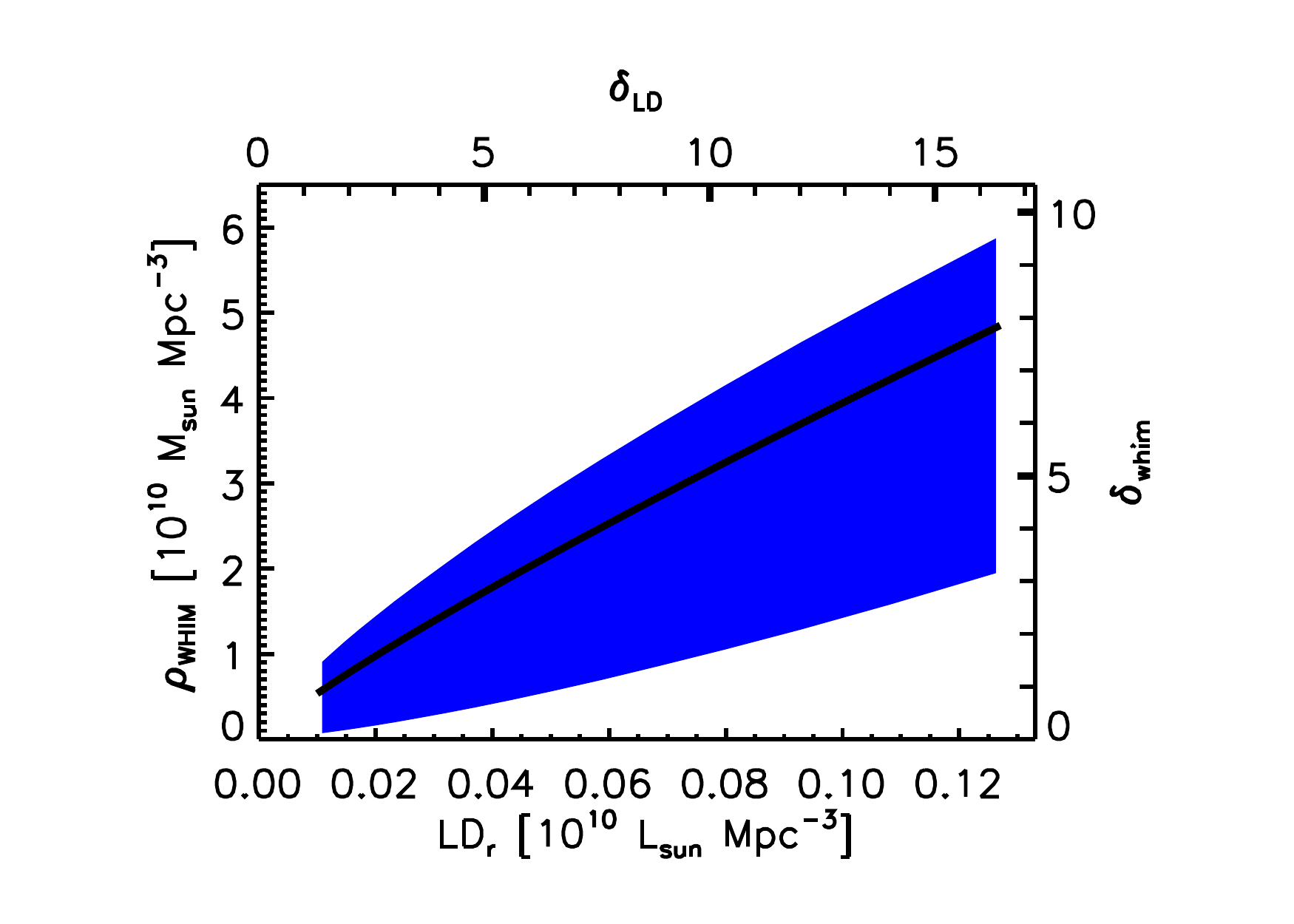} }
\vspace*{-0.5cm}
\caption{The power-law approximation of the relation between the luminosity density (LD$_r$) and the WHIM density ($\rho_{\rm whim}$) within the galaxy filaments 
in the simulations of \citetalias{2012MNRAS.423.2279C} (solid line), and the 1$\sigma$ uncertainty interval (blue region). 
Left panel shows the full fitted LD range while the right panel shows the $\delta_{\rm b}~\sim$1--10 range.}
\label{rel.fig}
\end{figure*}

The parameter uncertainties in Eqs.~\ref{rel1.eq} and \ref{rel2.eq} reflect the scatter among the power-law fits to the different Monte Carlo realisations, which is Gaussian when considering the logarithm of the quantities. We estimated the parameter uncertainties assuming that $A$ and $B$ are fully independent. This assumption is probably not fully valid and thus the parameter uncertainties are only approximate. While the normalisation parameter $A$ is formally constrained within 20\%, the correlation between $A$ and $B$ (the higher normalisation may be compensated by a the steeper slope to some extent) is expected to yield larger variations for the WHIM density at given LD. To verify this, we calculated the 68\% confidence intervals for the WHIM density at different LD values using the WHIM density predictions using the power-law model with a pair of $A$ and $B$ values from each power-law fit to the randomised data (i.e. including the parameter correlations). In the range $\delta_{\rm b}\sim$10--100 the WHIM density is uncertain by a factor of 2--3, while at the lower densities ($\delta_{\rm b}\sim$1--10) the variation reaches a factor of $\sim$10 (see Fig.~\ref{rel.fig}).

\section{Testing the method}
\label{sculptor}
In the current work our emphasis is on a detailed description of our WHIM method. In a follow-up paper we will carry out a full testing of the reliability of our method using all available data of sufficient quality. Here we perform a first test on the accuracy of our LD-based WHIM column density predictions by comparing them with those obtained independently via X-ray absorption measurements of a background blazar.

We chose to apply the first test to the Sculptor Wall~(SW) and Pisces-Cetus~(PC) superclusters, in the sight-line to the background blazar H2356-309, since they are 
1) among the closest reported X-ray absorption systems (z$\sim$0.03 and $\sim$0.06) respectively (\citetalias{2010ApJ...714.1715F} and \citetalias{2010ApJ...717...74Z})
2) covered with (2dF) galaxy survey data of sufficient quality (2dF, see \citealt{2001MNRAS.328.1039C, 2006AN....327..365T}) whose
3) reported WHIM column densities are among the highest in the literature.

\subsection{Luminosity density fields and galactic filaments}
\label{ldfield}
\begin{figure*}
	\centering
\includegraphics[width=16cm,angle=0]{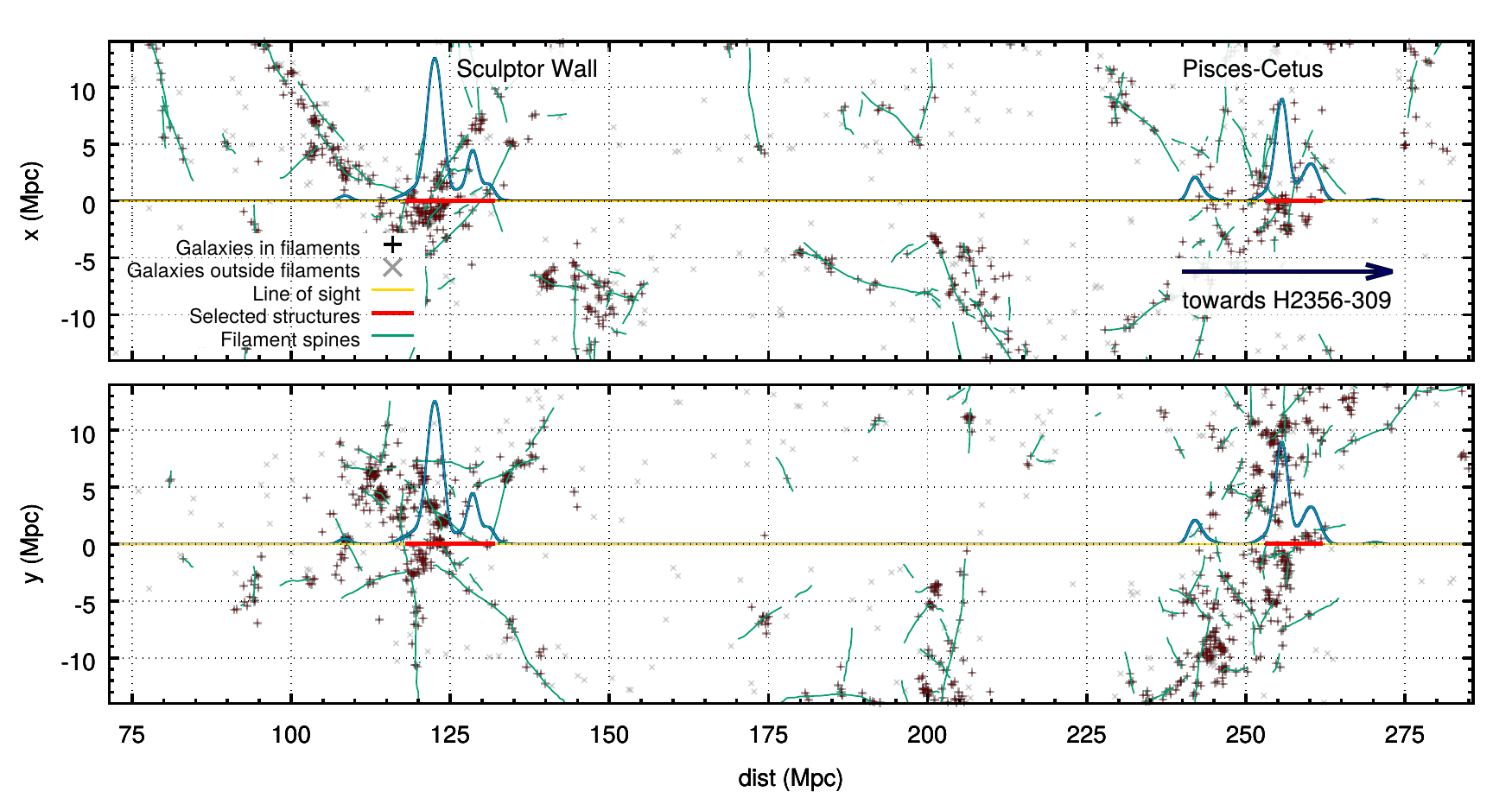}
\caption{The spatial distribution of 2dF galaxies around the H2356-309 sight-line (yellow line) projected at two orthogonal directions (upper and lower panels), at 
the distance range covering  Sculptor Wall and Pisces-Cetus structures.  The coordinates refer to the CMB rest frame. The galaxies belonging to filamentary 
environments are denoted with plus signs 
while other galaxies are denoted with crosses. The filament spines are indicated with green lines. The relative LD level at each radii is denoted with the blue 
line. The red parts of the sight-line highlight the radial ranges of the luminosity density profiles analysed in this work (see Fig.~\ref{LDprof.fig}).}
\label{sculp_gal.fig}
\end{figure*}

\begin{figure*}
	\centering
\includegraphics[width=16.0cm,angle=0]{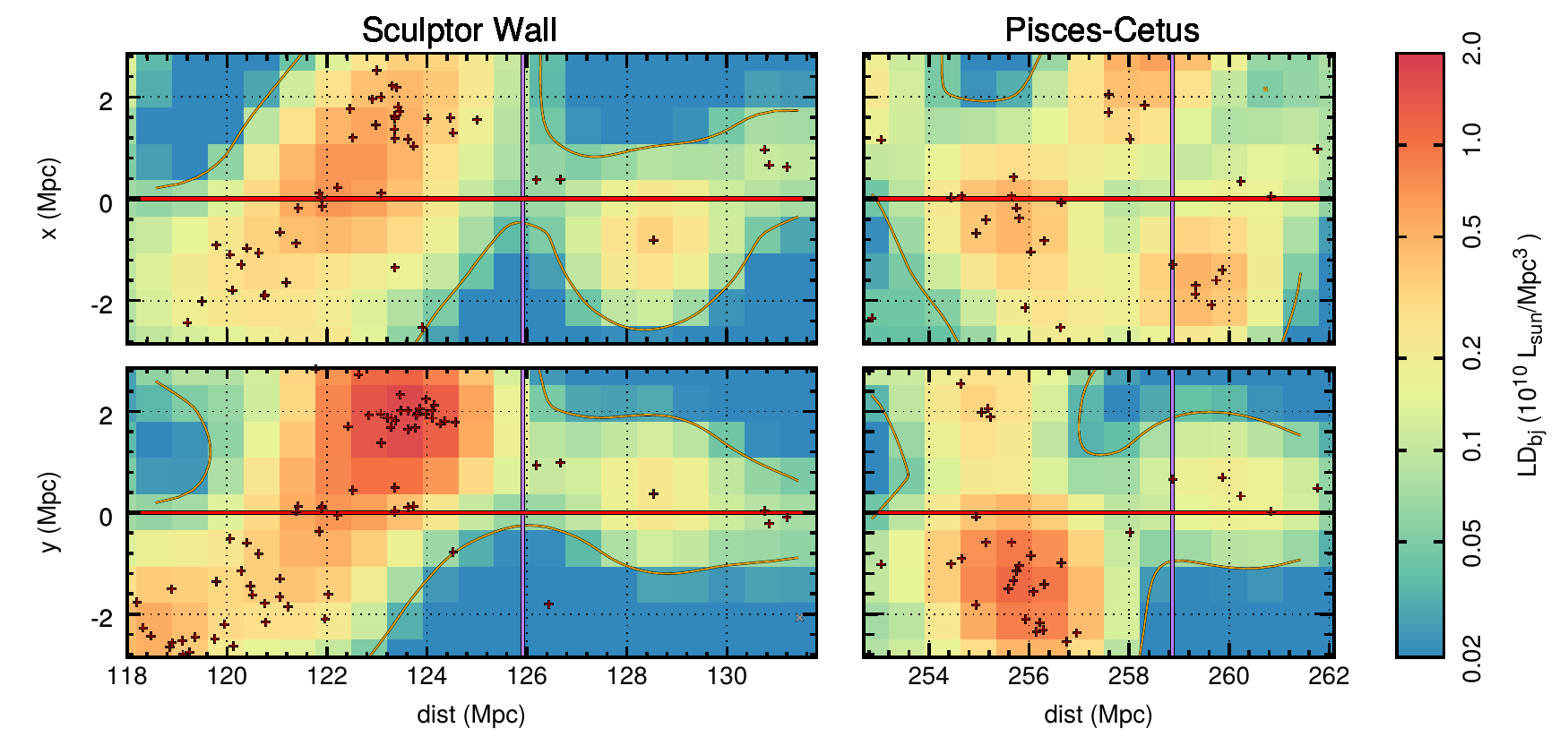}
\caption{The 2dF $b_j$ band luminosity density field slices of 1.4~Mpc thickness in two orthogonal directions (upper and lower panels) along the line-of-sight towards the blazar H2356-309 (red horizontal lines) in the Sculptor Wall (left panels) and Pisces-Cetus (right panels) structures. The contours indicate our adopted lower LD threshold LD$_{b_{j},\mathrm{min}} = 0.05\times 10^{10}L_{\odot}\,\mathrm{Mpc}^{-3}$ for extracting the data for further analysis. The radial ranges equal those used for the luminosity density profile analysis (see Fig.~\ref{LDprof.fig}). The galaxies relevant to the sight-line, i.e. within a distance of two times the smoothing kernel size from the sight-line (2.8~Mpc), are marked with dark red symbols: plus signs correspond to locations of galaxies belonging to a filamentary environment while other galaxies are denoted with crosses. The purple vertical lines indicate the centroids of the Chandra X-ray absorption lines (\citetalias{2010ApJ...714.1715F}; \citetalias{2010ApJ...717...74Z}).}
\label{sculp_ld.fig}
\end{figure*}

We applied the LD method (see Section~\ref{ld_meth}) to the $b_j$ band luminosities of the galaxies around SW and PC (see Fig.~\ref{sculp_gal.fig}) and thus produced the LD$_{b_j}$ fields (see Fig.~\ref{sculp_ld.fig}). Around the redshifts of the X-ray absorption line centroids of both SW and PC, the blazar sight-line passes through enhanced LD regions (see Fig.~\ref{sculp_ld.fig}). The enhancement of LD at the locations of the X-ray detected WHIM supports our finding from the simulations \citepalias{2012MNRAS.423.2279C} that LD traces the WHIM.

We then applied the Bisous model (see Section~\ref{bisous}) to the above galaxy distribution to identify galactic filaments. We found that there are several filaments within the enhanced LD regions (see Fig.~\ref{sculp_gal.fig}). There are 247 galaxies affecting the LD profiles of SW and PC, i.e. located within two times the size of the LD field smoothing kernel (2.8~Mpc) from the H2356-309 sight-line at the matching redshifts. Only two of these were classified as non-filament galaxies. This supports our finding from simulations \citepalias{2012MNRAS.423.2279C} that within the filamentary environments the WHIM density is enhanced from the cosmic average. Thus, it is justified to apply below the LD--WHIM density relation (Eq.~\ref{rel2.eq}) derived within the filamentary environments.

Since both absorbers have originally been targeted due to a priori knowledge of the large-scale concentrations of galaxies around them, the above agreement serves as a positive indicator of the performance of our filament-finder. We will perform a more robust test in a follow-up work.
 
The fortunate combination of the blazar position and the orientation of the filaments in both SW and PC results into relatively long ($\sim$10~Mpc) l.o.s. projection of high LD regions.
This contributes to these systems being among the most significantly X-ray detected WHIM structures to date.

\subsection{Luminosity density profiles and N$_H$}

\begin{figure*}
	\centering
\vspace{-10cm}
\hbox{
\hspace{-0.9cm}
\includegraphics[width=14.5cm,angle=0]{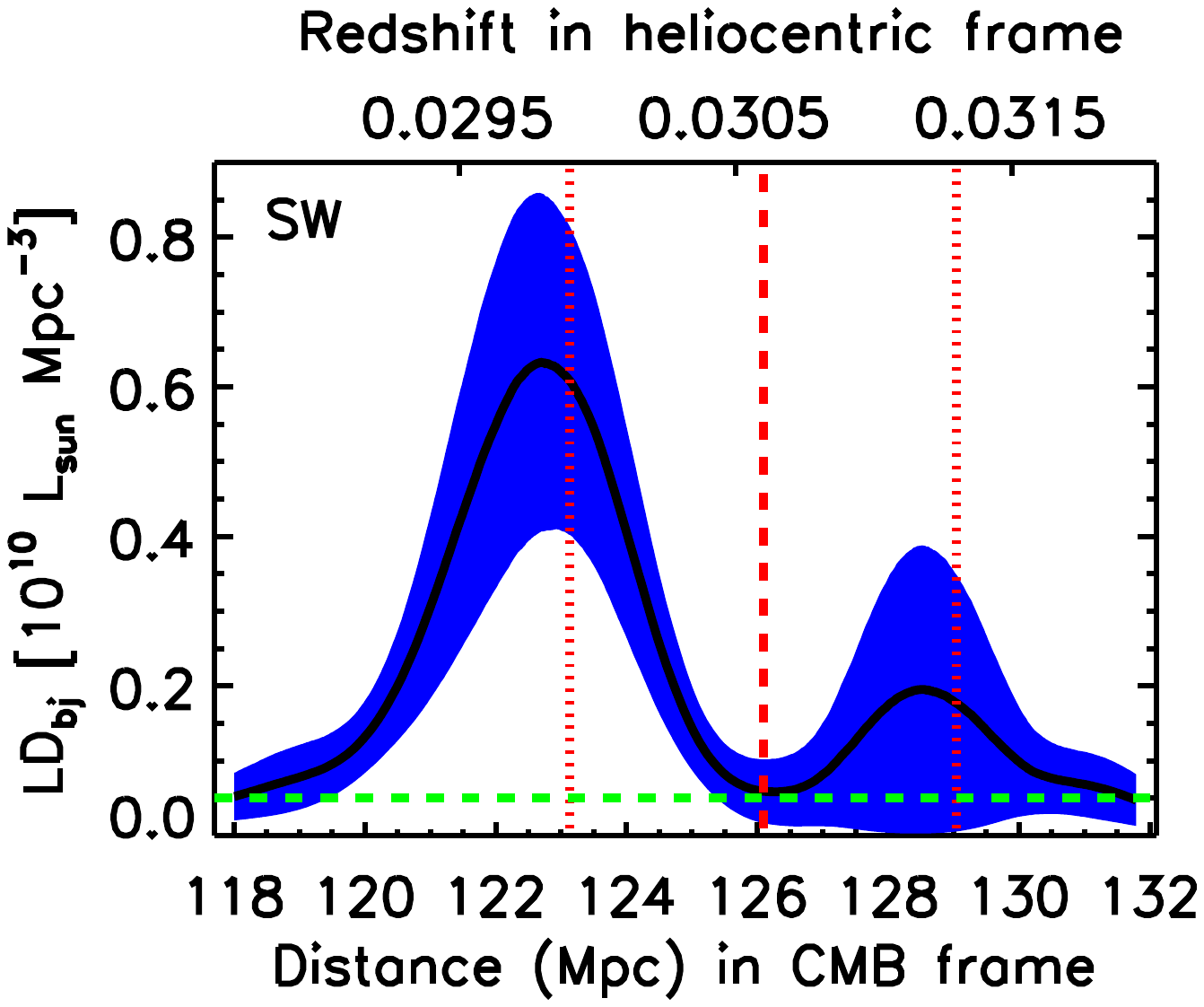}
\hspace{-5.5cm}
\includegraphics[width=14.5cm,angle=0]{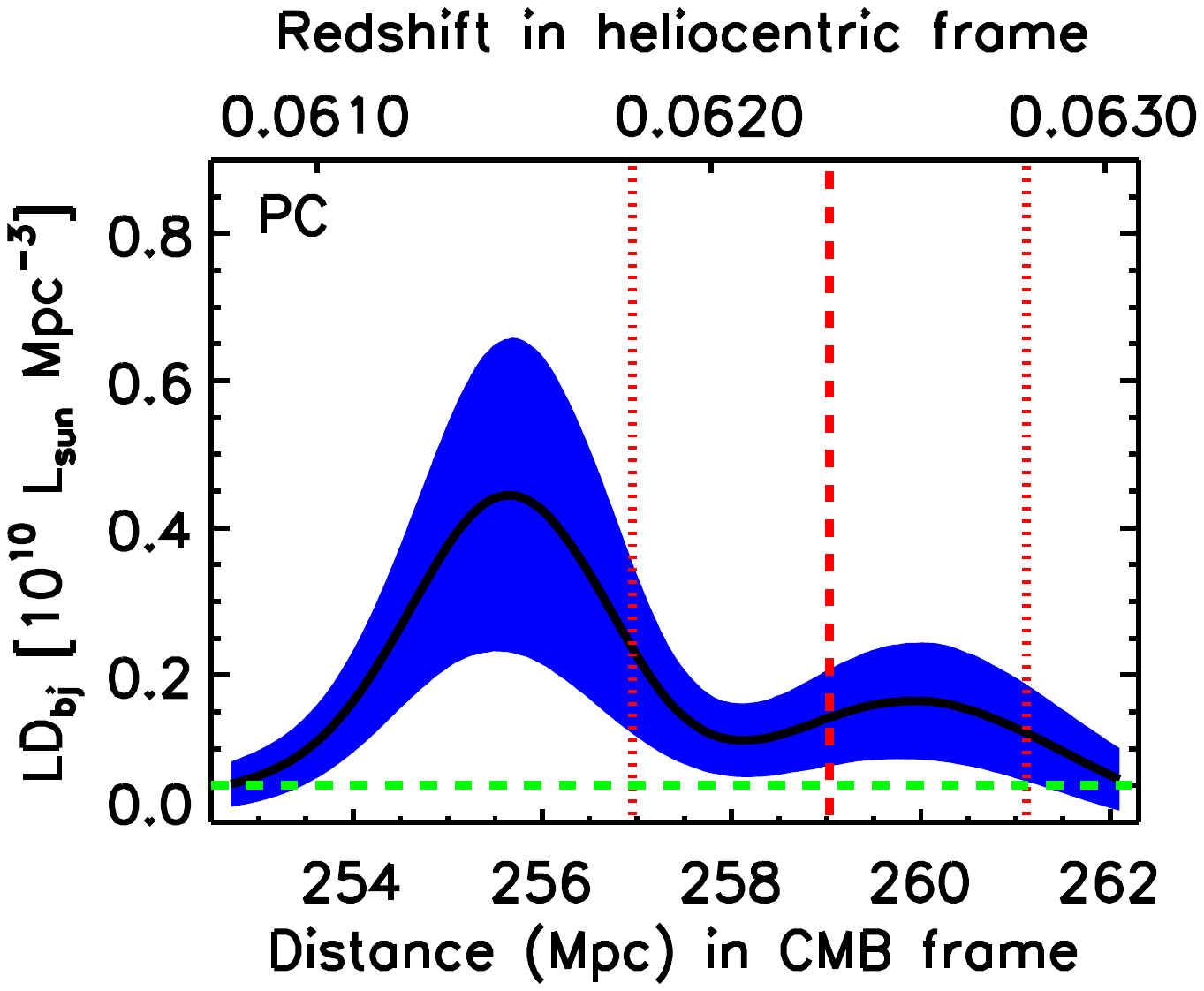}
}
\caption{The 2dF $b_j$ band luminosity density profiles of the Sculptor Wall (left panel) and Pisces-Cetus (right panel) along the H2356-309 sight-line
(black solid lines) together with 1$\sigma$ uncertainties (blue regions). The horizontal (green) dashed line indicates our adopted lower LD threshold 
LD$_{b_{j},min}$~=~0.05~$\times~10^{10}$~L$_{\odot}$~Mpc$^{-3}$ for extracting the data for further analysis.
The vertical dashed lines indicate the best-fit heliocentric redshifts of the Chandra X-ray lines while the dotted lines bracket the statistical 1 $\sigma$ uncertainties of 
the redshifts (from \citetalias{2010ApJ...714.1715F} and \citetalias{2010ApJ...717...74Z}). The co-moving distances refer to the CMB rest frame.}
\label{LDprof.fig}
\end{figure*}

In order to evaluate the column densities along the sight-line, we proceeded into characterising the radial behaviour of the luminosity density. We performed this by producing the LD$_{b_j}$ profiles of SW and PC by sampling the LD$_{b_j}$ fields along the line of sight towards the blazar H2356-309 (see Fig.~\ref{LDprof.fig}). By examining the LD fields and the LD profiles (see Figs.~\ref{sculp_ld.fig} and \ref{LDprof.fig}), we set a lower LD threshold LD$_{b_{j},\mathrm{min}} = 0.05\times 10^{10}L_{\odot}\,\mathrm{Mpc}^{-3}$ for considering the radial profile of the LD field across the X-ray line centroid and set the likely redshift range (z$_1$-z$_2$) of the absorber equal to that in which LD$_{b_{j}}~>~$LD$_{b_{j},\mathrm{min}}$.

Within the reported uncertainty range of the redshift of the X-ray absorber (\citetalias{2010ApJ...714.1715F}; \citetalias{2010ApJ...717...74Z}), there is a continuous, two-peaked structure of $\sim$10~Mpc length in both SW and PC (see Fig.~\ref{LDprof.fig}). 
The redshifts of the two peaks in the LD profiles are very similar.
Their separation, $\Delta_{z}~\sim~0.001$, corresponding to $\Delta_{\Lambda}~\sim~0.01$~\AA, at 20~\AA,
is below the energy resolution of both Chandra/LETGS and XMM-Newton/RGS.
Therefore we considered the full $\sim$10~Mpc long two-peaked structure as a single system for
both SW and PC.

We then converted our measured $b_j$ band Sculptor LD profiles into units of luminosity over-densities $\delta_\mathrm{LD}(z)=\mathrm{LD}_{b_j}(z)/\langle\mathrm{LD}_{b_j}\rangle$ (see the discussion in Section~\ref{galaxies}) by applying the mean $b_j$ band luminosity density in the full 2dF survey (see Table~\ref{conv.tab}).
We then used our LD--WHIM density relation (Eq.~\ref{rel2.eq}) to convert the above luminosity overdensity profiles into WHIM density profiles. Integration of the density profile then yielded the column density estimates using
\begin{equation}
N_H (\mathrm{WHIM}) = \int_{z_1}^{z_2} \rho_\mathrm{whim}(z)\, \mathrm{d}z. 
\label{int.eq}
\end{equation}

\subsection{Uncertainties of the estimated WHIM column densities}

In this Section we analyse the two main sources of uncertainties affecting the estimates of the WHIM column densities. First, we analyse the reliability of the observed luminosity density profiles. Second, we analyse the effect of the scatter in the calibration of the LD--WHIM density relation in simulations \citepalias{2012MNRAS.423.2279C}.

\begin{figure*}
	\centering
\vspace{-11cm}
\hbox{
\hspace{-0.9cm}
\includegraphics[width=14.5cm,angle=0]{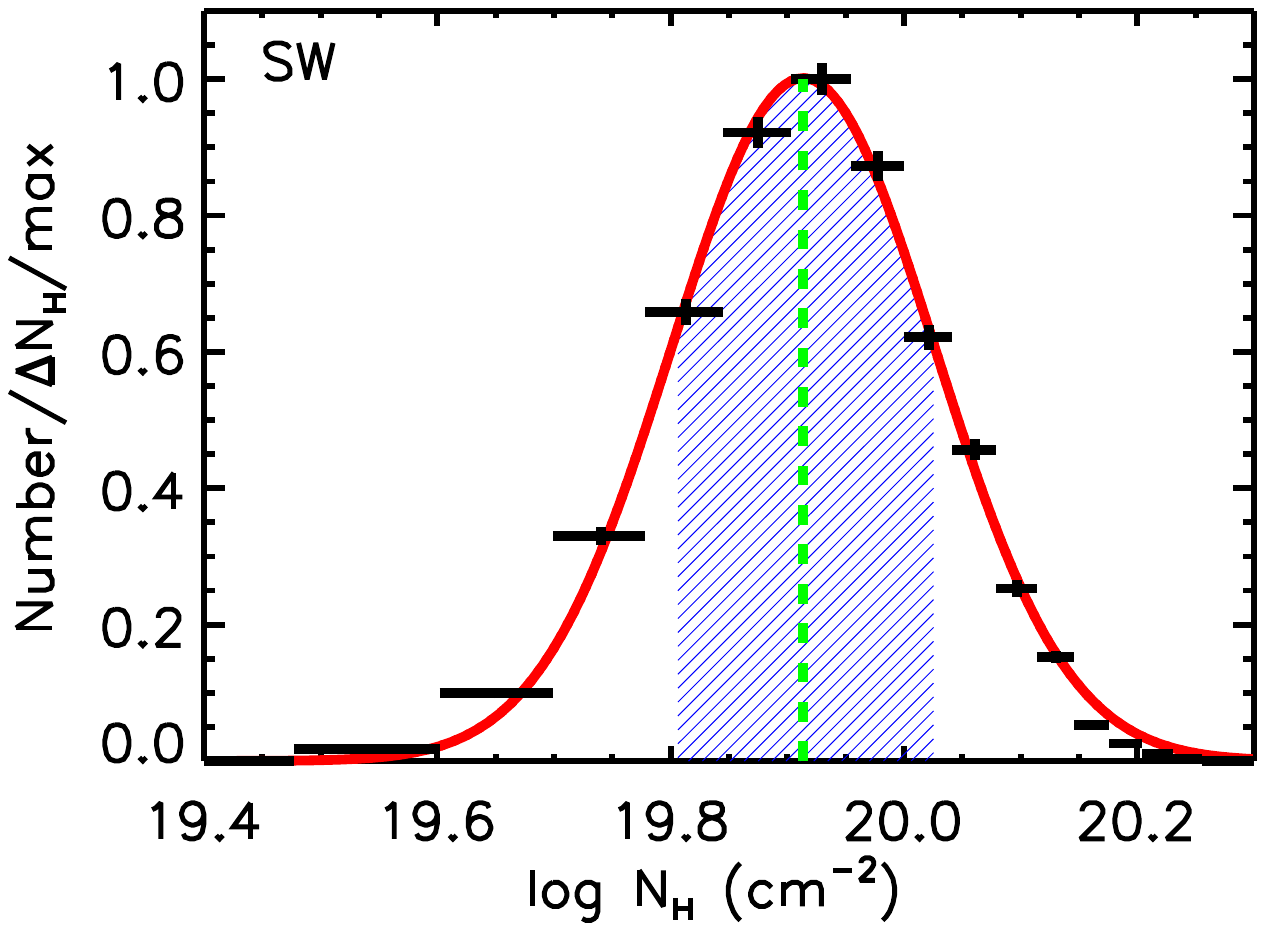}
\hspace{-5.5cm}
\includegraphics[width=14.5cm,angle=0]{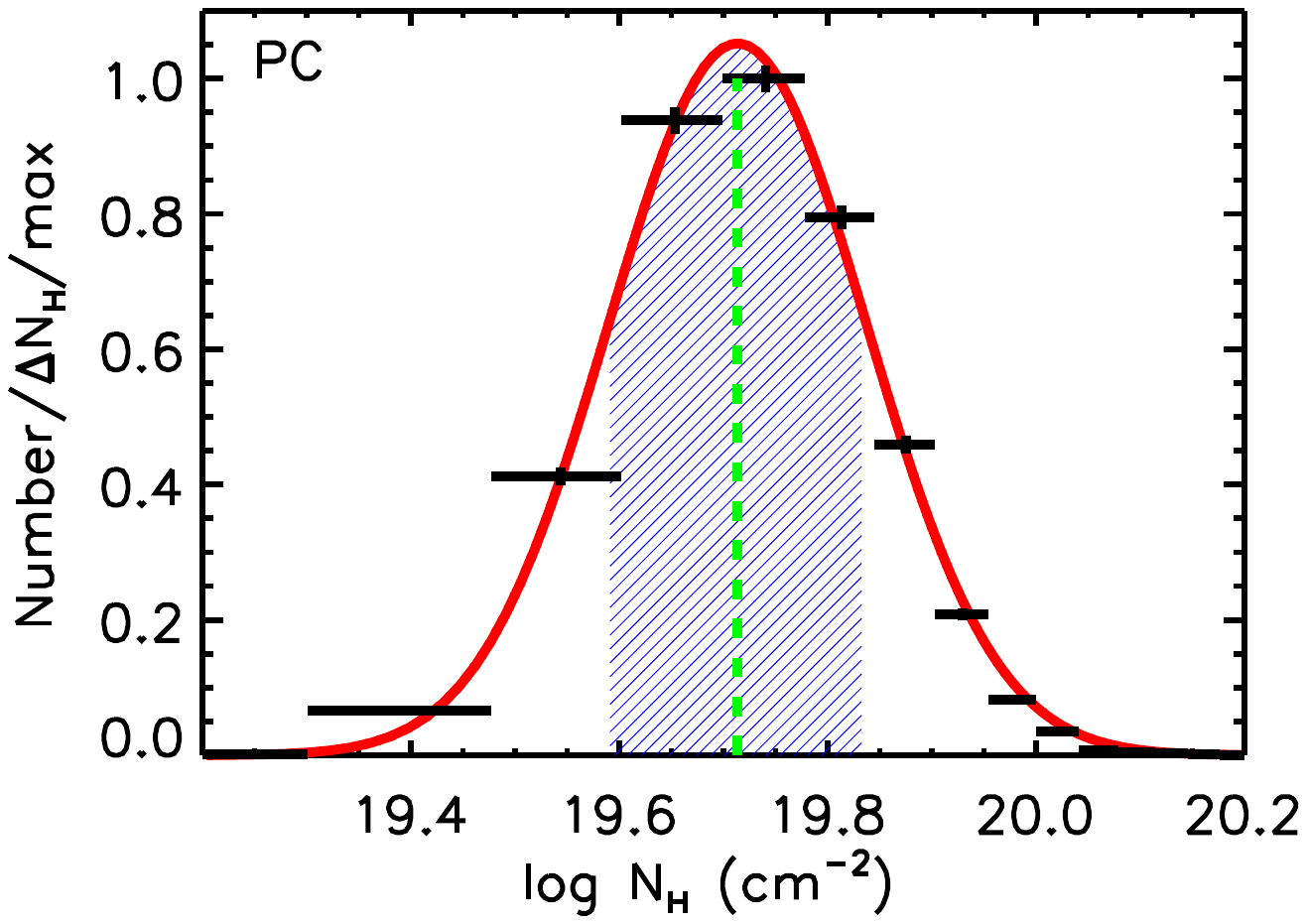}
}
\caption{
The WHIM hydrogen column density distributions (black crosses) due to the LD profile measurement uncertainties for the Sculptor Wall (left panel) and Pisces-Cetus (right panel) structures. The best-fit log-normal models are shown with (red) solid lines. The best value and the 1$\sigma$ confidence intervals are indicated with (green) dashed line and the (blue) shaded regions, respectively.}
\label{NHdist_simu.fig}
\end{figure*}

\begin{figure*}
	\centering
\vspace{-11cm}
\hbox{
\hspace{-0.9cm}
\includegraphics[width=14.5cm,angle=0]{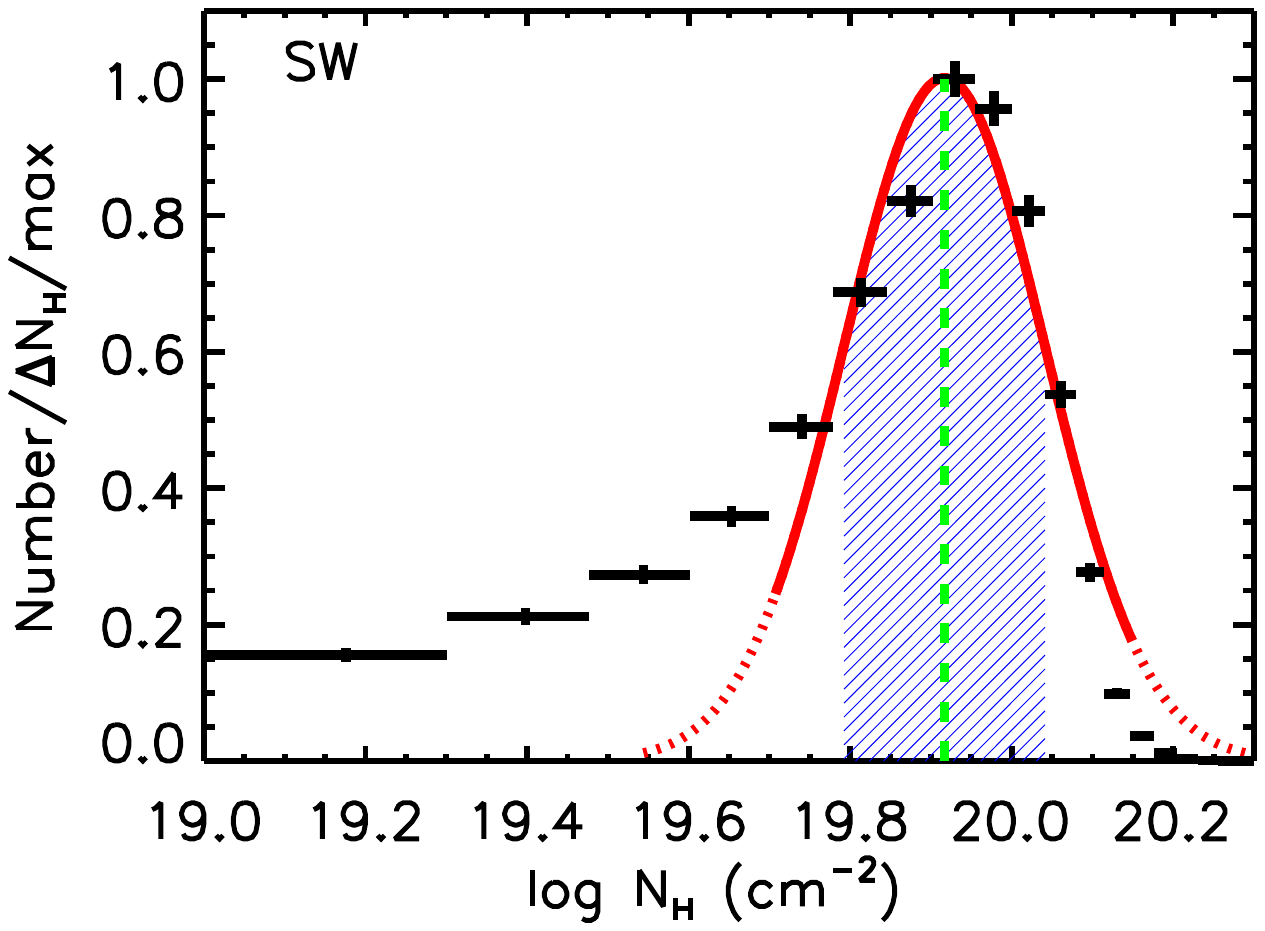}
\hspace{-5.5cm}
\includegraphics[width=14.5cm,angle=0]{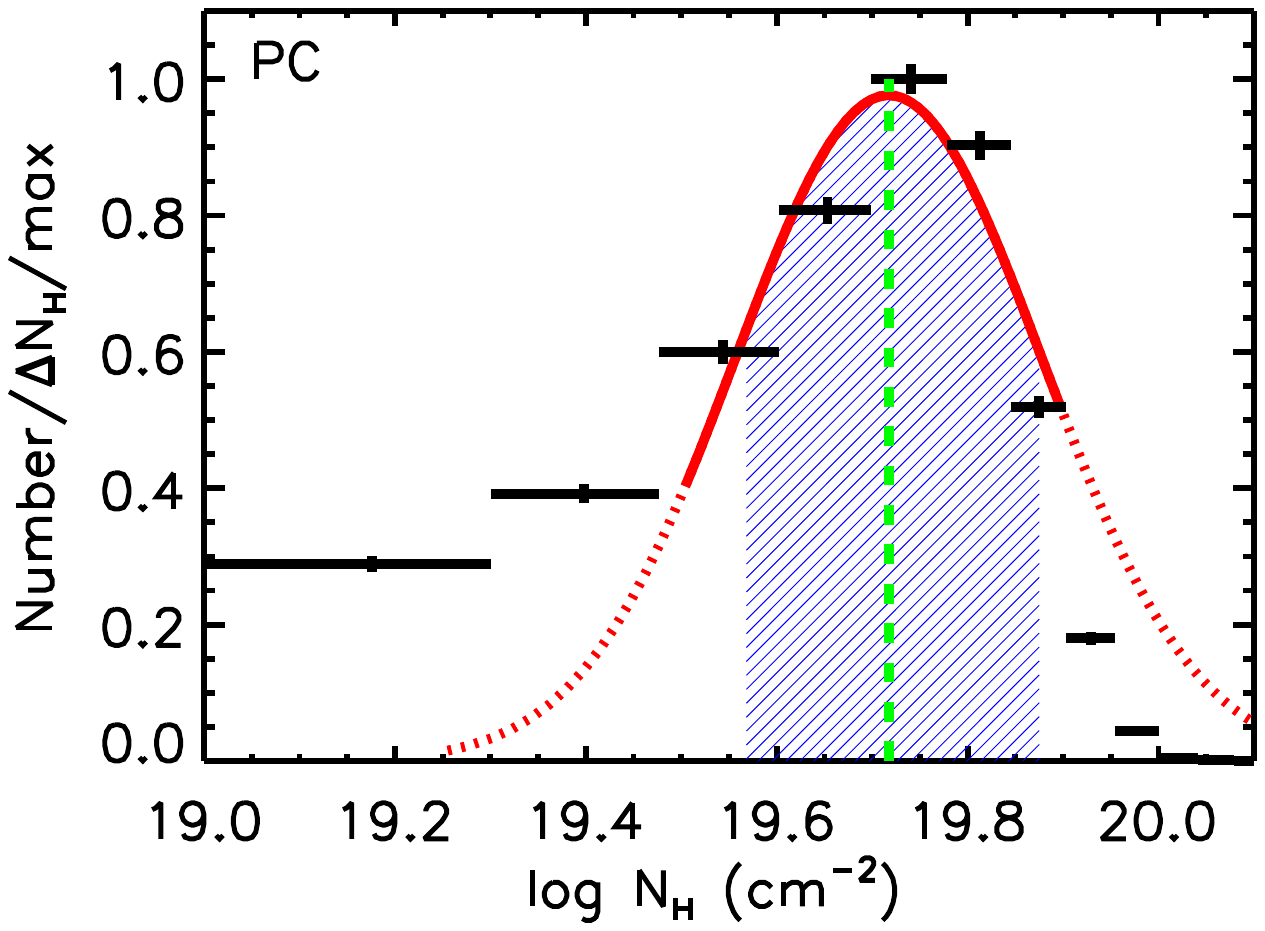}
}
\caption{The WHIM hydrogen column density distributions (black crosses) due to the LD--WHIM density scatter in the simulations \citepalias{2012MNRAS.423.2279C} for the Sculptor Wall (left panel) and Pisces-Cetus (right panel) structures. The best-fit log-normal models (red lines) in the fitted range (solid) are shown together with the extrapolation (dotted). The best value and the 1$\sigma$ confidence intervals are indicated with (green) dashed line and the blue shaded regions, respectively.}
\label{NHdist.fig}
\end{figure*}

\subsubsection{Uncertainties of the observed luminosity density field}
\label{bootstrap}

The main source of uncertainty in an observed LD profile is the shot noise due to the limited number of galaxies used for the LD field construction. We estimated this effect by applying a smoothed bootstrap procedure on the 2dF galaxy sample (see \citealt{2012A&A...539A..80L} for more details and justification of the use of the smoothed bootstrap in this context). The galaxies were selected with replacement as usual in bootstrap, but additionally their positions were randomised. The distance of the randomised position from the original one is defined by a smoothing kernel, whose size is half of that used for the original LD field smoothing in Section \ref{ld_meth}. Repeating this procedure 10\,000 times we obtained a set of bootstrapped LD profiles for SW and PC and we used the 68\% LD scatter interval at each radius to determine the uncertainties of the LD profile measurements (see Fig.~\ref{LDprof.fig}).

We propagated this scatter into N$_H$ measurement by converting each bootstrapped LD profile into a WHIM density profile using Eq.~\ref{rel2.eq} and integrating the profiles using Eq.~\ref{int.eq}. The resulting N$_H$ distributions followed well the log-normal prediction (see Fig.~\ref{NHdist_simu.fig}). We thus used the best-fit log-normal centroid and width parameters to obtain the best values and 1$\sigma$ intervals of N$_H$ i.e. 0.8~[0.6--1.1]~$\times~10^{20}$~cm$^{-2}$ and 0.5~[0.4--0.7]~$\times~10^{20}$~cm$^{-2}$ for SW and PC, respectively.

\subsubsection{Effect of scatter in the LD--WHIM density relation}
\label{errprop}
Due to the significant scatter of the WHIM density values for a given LD value in the cosmological simulations used in this work, our model prediction is statistical by nature. The following list contains brief discussion about the possible origins of the scatter.

\begin{itemize}

\item
The galaxy luminosity and the WHIM density do not solely depend on the underlying dark matter 
density but may depend on e.g. the stellar feedback mechanisms and the gas physics.

\item
The HOD used to populate the dark matter halos with galaxies is statistical by nature.

\item
The dark matter halo geometry may deviate from the assumed spherical symmetry in the HOD modelling.

\end{itemize}

We already characterised this scatter in Section~\ref{rel.sec} by the power-law fits to the set of randomised LD--WHIM density relations, when deriving the uncertainties of the LD--WHIM density power-law model parameters. Due to the parameter correlations (see Section~\ref{rel.sec}) we did not perform a simple analytical error propagation using Eqs.~\ref{rel2.eq} and~\ref{int.eq}. Rather, we used here the pairs of the best-fit power-law normalisation and index parameters obtained in Section~\ref{rel.sec} to obtain a set of LD--WHIM density relations which reflects the scatter of the WHIM density values for a given LD in \citetalias{2012MNRAS.423.2279C} 
simulations. We applied each of these scattered LD--WHIM density relations to the measured LD profiles of SW and PC first to obtain a set of WHIM density profiles and then, by integration, we obtained distributions of N$_H$ values for both structures (see Fig.~\ref{NHdist.fig}).

At N$_{H}$ values within the 68\% confidence interval around the centroid, the distributions can be reasonably well approximated with a log-normal model. We thus used the best-fit log-normal centroid and width parameters to obtain the best values and 1$\sigma$ intervals of N$_H$ i.e. 0.8~[0.6--1.1]~$\times~10^{20}$~cm$^{-2}$ and 0.5~[0.4--0.8]~$\times~10^{20}$~cm$^{-2}$ for SW and PC, respectively.
At the lowest and the highest N$_{H}$ values the distribution significantly exceeds the log-normal distribution. The low end tail is a consequence of the high end tail of the index distribution (see Fig.~\ref{norm_index.fig}).
Note that the rather large (by a factor of $\sim$10) uncertainty of the LD--WHIM density relation at the lowest densities (see Fig.~\ref{rel.fig}) has a negligible effect in case of SW and PC, since most of the measured LD values are much higher (see Fig.~\ref{LDprof.fig}).

\begin{table*}
 \centering
  \caption{Properties of the absorbers.}
\label{NH.tab}
  \begin{tabular}{lcccc}
  \hline\hline
    & X-ray  & LD & X-ray & LD \\   
\hline\hline
                         &                                     &                                      &                &                       \\
                         & \multicolumn{2}{c}{Sculptor Wall}                                      &  \multicolumn{2}{c}{Pisces-Cetus} \\
                         &                                     &                                      &                &                       \\
$z$\tablefootmark{a}       & 0.0310--0.0334                      &  0.0286--0.0320                      & 0.0618--0.0628 & 0.0608--0.0630          \\
N$_H$ ($10^{20}$cm$^{-2}$) &     --                              &  0.8$\pm$0.4                         &   --           & 0.5$\pm$0.3           \\
$\log{\mathrm{N}_H}$              &    --                               &  19.9$^{+0.1}_{-0.3}$                   & 20.1$\pm$0.2\tablefootmark{b}  & 19.7$^{+0.2}_{-0.3}$    \\
$\log{\mathrm{N}_{\ion{O}{VII}}}$\tablefootmark{c}          & 16.8$^{+1.3}_{-0.9}$ &  15.7$^{+0.2}_{-0.3}$  & --            & --                   \\
\hline
\end{tabular}
\tablefoot{
\tablefoottext{a}{{\it X-ray:} The uncertainty interval of the {\it Chandra} X-ray absorption line redshift 
in Sculptor Wall \citepalias{2010ApJ...714.1715F} and Pisces-Cetus \citepalias{2010ApJ...717...74Z},
{\it LD:}  The extent ($z_1$--$z_2$) of the LD structure intersected by the H2356-309 sight-line. The redshifts refer to the heliocentric frame.}
\tablefoottext{b}{{The more significant warm component}}
\tablefoottext{c}{Derived from N$_H$ assuming $T = 10^6$~K,  O abundance of 0.1~Solar, and Solar O/H ratio from \citet{1998SSRv...85..161G}.}
}
\end{table*}

\subsection{Final N$_H$ values}
The effects of 1) the Sculptor LD profile measurement uncertainties using the 2dF data and 2) the LD--WHIM density scatter in the cosmological simulations,
on the N$_H$ estimates are similar and significant (i.e. a variation by a factor of $\sim$2). We thus combined the two uncertainties in quadrature. 
The best values differ slightly and thus we averaged those, obtaining the final results that are given in Table~\ref{NH.tab}.

\subsection{Comparison with X-rays}

Our procedure predicts the hydrogen column density of the WHIM from the galaxy luminosity density. However, the typical signatures of the WHIM in the X-ray band are line features from highly ionised metals, like \ion{O}{VII}. The conversion between the hydrogen column density and metal ion column density depends on the ionisation fraction (i.e. the temperature) and the metal abundance. If only a single absorption line is measured, as in the case of the Sculptor Wall structure (\citetalias{2010ApJ...714.1715F}), the temperature and column density constrains are very poor. If we assume that oxygen abundance is 0.1~Solar and that the temperature is $10^6$~K (i.e. the ionisation fraction of \ion{O}{VII} is 1.0) our N$_H$ estimate for the SW structure ($\log{N_H} = 19.9^{+0.1}_{-0.3}$) translates to $\log{N_{\ion{O}{VII}}}~=~15.7^{+0.2}_{-0.3}$, consistent within the large uncertainties of the X-ray measurement of \citetalias{2010ApJ...714.1715F}, $\log{N_{\ion{O}{VII}}}~=~16.8^{+1.3}_{-0.9}$ (see Table~\ref{NH.tab}).

In the case of the Pisces-Cetus structure, lines from several elements and ionisation stages were reported \citepalias{2010ApJ...717...74Z} for a warm absorber\footnote{\citetalias{2010ApJ...717...74Z} also reported a possible hotter component at the same redshifts but due to its lower significance we do not consider it here.} ($\log{T}~(\mathrm{K})\sim~5.4$). Thus the temperature and the equivalent hydrogen column density were well constrained and we can make a direct comparison (assuming that oxygen abundance is 0.1~Solar). Our estimate ($\log{N_H}~=~19.7^{+0.2}_{-0.3}$) is consistent within 1$\sigma$ uncertainties with that of \citetalias{2010ApJ...717...74Z}, i.e. $\log{N_H} = 20.1\pm0.2$.

The agreement of our LD-based prediction for the column density of the WHIM in SW and PC structures with that obtained from X-ray absorption measurements indicates that our method is robust. This also indicates that the luminosity density and the galactic filaments are reliable signposts for the WHIM.

\section{Galaxy confusion}

In the above we have interpreted the X-ray absorption features being due to warm-hot {\it intergalactic} medium. We examine here an alternative hypothesis, the galaxy confusion, which states that instead of the intergalactic diffuse WHIM, the absorption is caused by the X-ray halos of accidental galaxies close to the studied sight-line \citep[e.g.][]{2013ApJ...762L..10W}. 
Note that in this case the match of our intergalactic LD - based WHIM column densities and the X-ray ones  (\citetalias{2010ApJ...714.1715F}; \citetalias{2010ApJ...717...74Z}) is not causal, but rather a co-incidence.

\begin{table*}
 \centering
  \caption{Nearby galaxies
  \label{gal.tab}}
  \begin{tabular}{lccccccl}
  \hline
 \hline
 ID     & RA         & DEC       & $z$\tablefootmark{a} & $L_{b_j}$             &  $M_{b_j}$    & $d$\tablefootmark{b}   & type\tablefootmark{c}      \\
        &    deg     & deg       &                    & $10^{10}\,L_{\odot}$  &              & kpc                  &                            \\
 \hline
 \hline
        &            &           &                                   &                      &            &       &       \\
        & \multicolumn{7}{c}{ Sculptor Wall} \\
        &            &           &                                   &                      &            &       &       \\

1       & 359.79297  & $-30.58761$ & 0.0296                            & 0.08                 & $-17.3$      & 90   &  star forming       \\
2       & 359.70248  & $-30.56833$ & 0.0297                            & 0.04                 & $-16.7$      & 190  &  star forming       \\
3       & 359.90756  & $-30.65583$ & 0.0301                            & 0.62                 & $-19.5$      & 240  &  elliptical         \\
4       & 359.84072  & $-30.80313$ & 0.0295                            & 0.29                 & $-18.7$      & 390  &  low star formation \\
5       & 0.12302    & $-30.72346$ & 0.0316                            & 0.08                 & $-17.3$      & 700  &  star forming       \\
6       & 359.30189  & $-30.61506$ & 0.0295                            & 0.02                 & $-15.8$      & 880  &  low star formation \\
7       & 359.36436  & $-30.46063$ & 0.0312                            & 0.88                 & $-19.9$      & 890  &  low star formation \\
8       & 0.27432    & $-30.61173$ & 0.0316                            & 0.06                 & $-16.9$      & 970  &  star forming       \\
        &            &           &                                   &                      &            &      &                    \\
        &  \multicolumn{7}{c}{Pisces-Cetus} \\
        &            &           &                                   &                      &            &      &                    \\
9       & 359.79467  & $-30.62261$ & 0.0627                            & 0.14                 & $-17.9$      & 50   &  active starburst   \\
10      & 359.87066  & $-30.55599$ & 0.0625                            & 0.21                 & $-18.4$      & 470  &  active starburst   \\
11      & 359.79755  & $-30.76000$ & 0.0613                            & 0.13                 & $-17.8$      & 590  &  star forming       \\
12      & 359.60680  & $-30.64623$ & 0.0612                            & 0.09                 & $-17.4$      & 680  &  low star formation \\
13      & 359.76432  & $-30.84869$ & 0.0623                            & 0.11                 & $-17.6$      & 990  &  low star formation \\
        &            &           &                                   &                      &            &      &                    \\
\hline
\end{tabular}
\tablefoot{
\tablefoottext{a}{Redshifts are reported in the heliocentric frame.}
\tablefoottext{b}{distance from the H2356-309 sight-line in the plane of the sky.}
\tablefoottext{c}{Based on \citet{2002MNRAS.333..133M}.}
}
\end{table*}

The sole existence of a galaxy within a virial radius from an X-ray absorber is obviously no proof that the galaxy and its circum-galactic gas are the origin of the X-ray absorption features that we regard as WHIM signatures. We explored the galaxy confusion hypothesis quantitatively in this work by 1) estimating the density of the X-ray halos of the nearby galaxies at the distance of H2356-309 sight-line and 2) comparing these estimates with the X-ray measurements. As possible candidates we very generously considered all such 2dF galaxies within the redshift range matching those of the SW and PC structures, which are located within 1 Mpc (i.e. several times the virial radii) from the H2356-309 sight-line. There are eight (five) of such candidates for SW (PC), (see Table~\ref{gal.tab}).

In the above list there are no massive galaxies which could be obvious candidates for the galaxy confusion. 
It is possible that close to the sight-line there are galaxies fainter than the 2dF limit, corresponding to $M_{B}~\sim -16$ at these redshifts.
However, these luminosities correspond to $\sim$1\% of that of the Milky Way. Thus, one needs 100 of such objects to reach the mass of the Milky Way
and possibly a similar WHIM N$_{H}$ level of $10^{19}$~cm$^{-3}$ (assuming the temperatures of such haloes do reach WHIM values).
These objects should be clustered at the sight-line, at the redshifts consistent with the X-ray measurements. Thus these objects should be more clustered 
than Milky Way - like galaxies (see Table~\ref{gal.tab}), i.e. their bias parameter should be larger. This is inconsistent with observations and simulations that 
show that faint galaxies are less clustered than the luminous ones (e.g. \citet{2002MNRAS.332..827N}; \citetalias{2012MNRAS.423.2279C}).

Galaxy~3 in our list is the only elliptical galaxy while all others are spirals. In the following we examine separately the two galaxy types. For the comparison with other works on galaxies, we approximated the absolute $b_j$ band magnitudes and luminosities with those in the $B$-band \citep{2002MNRAS.336..907N}.

\subsection{Elliptical galaxy 3}

\citet{2013ApJ...762L..10W} suggested that the elliptical galaxy~3 in our list, at a distance of 240~kpc from the H2356-309 sight-line at the redshift of the Sculptor Wall, may be responsible for the X-ray absorption measured by \citetalias{2010ApJ...714.1715F}. 
The $b_j$ band magnitude of the galaxy corresponds to L$_{B}~\sim~10^{10}$~L$_{B \odot}$.  
From the analysis of X-ray halos of elliptical galaxies  \citep{2006ApJ...636..698F}, the corresponding 
X-ray luminosity of this object would be $L_{X}~\sim~10^{40}$~erg~s$^{-1}$ and a temperature of  T$\sim~10^{6}$~K, consistent with the WHIM properties.
Extrapolating the corresponding gas density profile in  \citep{2006ApJ...636..698F} to
a radius of 240~kpc (corresponding to the impact parameter of the sight-line to H2356-309) yields a density level of $\sim~10^{-5}$~cm$^{-3}$.
With such a low density a path length of $\sim~10^{25}$~cm, i.e. $\sim$10~Mpc would be required to build up the measured level of the column density, i.e.
N$_{H}~\sim~10^{20}$~cm$^{-2}$. This is clearly inconsistent with the virial radius of the galaxy, estimated as 350~kpc in \citet{2013ApJ...762L..10W}.

\subsection{Spirals}

All the spiral galaxies in the candidate list have a much smaller or similar $B$-band luminosity $L_B$ compared to the Milky Way, whose $M_{B}~\approx~M_{b_j}\sim -20$ \citep{1998gaas.book.....B, 1999AJ....118..883P, 2000PASP..112..529V}. We wish to use the L$_B$ values for comparing the stellar masses of the candidate spirals with that of the Milky Way. This is complicated since the $B$-band may be ``contaminated'' due to star formation. All our candidate spirals are star-forming and thus they likely have a similar bias in the $B$-band - based mass estimates of the candidates and the Milky Way. However, in order to form a conservative estimate 
we assumed a factor of 100 scatter in the mass/$L_B$ ratio, rendering the most massive candidate spirals 100 times more massive than the Milky Way.

The intensity of the X-ray emission from $T\sim~10^{6}$~K gas in several edge-on spiral galaxies, e.g. NGC~5775 \citep{2008MNRAS.390...59L} and NGC~4631 \citep{2009PASJ...61S.291Y}, has been observed to decrease exponentially with the height from the galactic plane. The joint X-ray emission and absorption measurements of \ion{O}{VII} and \ion{O}{VIII} lines of the Milky Way halo have obtained temperatures $\sim$$10^6$~K (consistent with WHIM) and equivalent hydrogen column densities at $10^{19}$~cm$^{-2}$ level \citep[e.g.][]{2010PASJ...62..723H, 2014PASJ...66...83S}, i.e. $\sim$10\% of that consistently determined by us and via X-rays in Sculptor Wall and Pisces-Cetus. The resulting path length scale of the absorber and thus the extent of the WHIM halo in spiral galaxies in the above works is consistently constrained below 10~kpc. While this indicates that the halo is identified with the stellar component, this also conflicts with the reports of the Galactic $\sim$100~kpc WHIM halo \citep[e.g.][]{2012ApJ...756L...8G}. In fact, \citet{2012arXiv1211.4834W} reported numerous significant errors in \citet{2012ApJ...756L...8G} which resulted in an order of magnitude overestimation of the Galactic WHIM halo size.

Since our candidate spirals have masses less than 100 times that of the Milky Way (see above), we assume here conservatively that all our candidate spirals have WHIM halos with a similar exponentially decreasing density profile as that of the Milky Way, except that scaled up by a factor of 100, i.e. with a central density of 
$10^{-1}$~cm$^{-2}$ and a scale height of 3~kpc. The most likely candidate for galaxy confusion is galaxy~9 (see Table~\ref{gal.tab}), the nearest one to the absorbers. \citet{2013ApJ...762L..10W} associated this galaxy with the X-ray absorber at Pisces-Cetus \citepalias{2010ApJ...717...74Z}. Extrapolating the exponential density profile to 50~kpc (the distance between the galaxy and H2356-309 sight-line), we obtain a very low WHIM density of $10^{-8}$~cm$^{-3}$. Let's assume that the column density of $\sim$$10^{20}$~cm$^{-2}$ is achieved along an absorber path length of 10~kpc ($\sim 10^{22}$~cm). Consequently, the required number density of the WHIM halo is $\sim 10^{-2}$~cm$^{-3}$ at the distance of the H2356-309 sight-line from the centre of a given galaxy. Thus, the exponential density profile of galaxy~9 fails totally to produce the measured level of WHIM column density. Since the other galaxies are more distant, the discrepancy is even bigger for those. Unless we underestimated the candidate spiral to Milky Way mass ratio by a factor of $10^6$ due to the star formation issue, the candidate spirals cannot produce the measured absorption in Pisces-Cetus.

In general, considering the projection effects, a Milky Way - like galaxy must be located within $\sim$10~kpc to the blazar sight-line in order to produce WHIM column densities at a level of $\sim 10^{19}$~cm$^{-2}$.

\section{Discussion}
In this work we tested the ability of our method to infer the presence of the WHIM by targeting large scale structures associated to putative WHIM 
detections. In the future we shall apply our method to perform a blind WHIM search along the sight-lines to distant quasars located within the areas of
the 2dF and SDSS galaxy surveys.
Note that our method is not limited to the directions towards the brightest background blazars, as is the case for the current rare WHIM detections based on the identification of characteristic X-ray and far ultra violet absorption lines. Rather, our method has potential for extending the hunt for the missing baryons to the full volumes of the spectroscopic galaxy surveys like SDSS and 2dF.

Our method can be used to estimate the mass density of the WHIM. Combined with the independent X-ray or FUV absorption measurements of the same WHIM structures, our additional constrain has potential for improving the diagnostics of the WHIM physics. In particular, there is a possibility of breaking the degeneracy between the WHIM column density and the abundance of the absorbing metal, and thus providing constraints for the latter. This could improve our understanding of the chemical and thermal enrichment of the intergalactic medium via supernovae explosions and AGN feedback. Currently these effects are poorly known and modelled in the hydrodynamical simulations.

Our LD--WHIM density relation together with the filament-tracing technique thus has potential for mapping a large sample of WHIM properties with the currently existing data. This is necessary for properly assessing the contribution of WHIM to the cosmological mass density budget. Thus our method has potential for advancement in the cosmological missing baryons problem in the near future. Our methods and the potential WHIM maps could be used to optimise the observational strategies for the search of the missing baryons with next generation satellites like Athena.

\section{Conclusions}

In this work we proposed, implemented, applied and tested a novel method to identify the elusive WHIM using galaxies in the filaments of the cosmic web. Our method rests upon a correlation between the density of the ``invisible'' WHIM and that of the observed galaxy luminosity. The calibration of the method was performed using the hydrodynamical numerical simulation \citepalias{2012MNRAS.423.2279C} and HOD-based mock galaxy catalogues mimicking the properties of SDSS galaxies, following the \citet{2011ApJ...736...59Z} prescriptions. To test the performance of our method on observational data we applied it to the distribution of galaxies in the 2dF survey, focusing on the Sculptor Wall and Pisces-Cetus superclusters. The main results, obtained both in the calibration and application phases are summarised below.

\begin{itemize}

\item
In the \citetalias{2012MNRAS.423.2279C} simulation the WHIM is preferentially associated to the filamentary structures. We identified the latter by applying the Bisous model \citep{Stoica:05, 2014MNRAS.438.3465T} to the distribution of the mock galaxies and found that the mass fraction of the gas in WHIM phase  
associated to these structures is $\sim$70\%, by a factor of $\sim$1.6 higher than the average in the full simulation box.

\item
When considering the WHIM gas and the galaxy luminosity in the filament environments, we found a significant correlation between their densities. The Pearson correlation coefficient between the two turned out to be $\sim$80\% and the relation between the WHIM gas overdensity and the galaxy luminosity overdensity consistent with linear: $\delta_{\rm whim}~=~0.7~\pm~0.1~\times~{\delta_\mathrm{LD}}^{0.9 \pm 0.2}$. This suggests that both the diffuse gas in the WHIM and the stellar component in the galaxies trace the same underlying dark matter density field.

\item
The above relation can be used in reverse, i.e. given the observed luminosity density in the filamentary structures identified in the spatial distribution of 
galaxies in galaxy redshift surveys one can use the relation to infer the column density of the WHIM.

\item
When testing our method on observational data, we found that the luminosity density was significantly enhanced from the average at the locations of independently detected WHIM (\citetalias{2010ApJ...714.1715F}; \citetalias{2010ApJ...717...74Z}). This indicates that the luminosity density traces the WHIM, consistently with what we found in the simulations.

\item
Our LD-based predictions for the column density of the WHIM in the Sculptor Wall and Pisces-Cetus superclusters agree with those obtained via X-ray absorption. This agreement indicates that our method is robust in estimating the density of the WHIM. Also, the galaxy filaments and the luminosity density are reliable signposts of the WHIM.

\item
The fortunate combination of the {\it angular position} of the {\it bright} background blazar H2356-309 behind {\it suitably oriented}, {\it long} filaments in the Sculptor Wall and Pisces-Cetus superclusters results into relatively long ($\sim$10~Mpc) line-of-sight projection of high LD regions. This contributes to these systems being among the most significantly X-ray detected WHIM structures to date.

\item
The signal that we detected cannot originate from the halos of the nearby galaxies since they cannot account for the large WHIM column densities that our method and X-ray analysis consistently find in the Sculptor Wall and Pisces-Cetus superclusters.

\end{itemize}


\begin{acknowledgements}
JN is funded by PUT246 grant from Estonian Research Council. We acknowledge the funding from the ESF grants IUT26-2, IUT40-2 and TK120 (the European Structural Funds grant for the Centre of Excellence ``Dark
Matter in (Astro) particle Physics and Cosmology'').
Thanks to F.~Pace for providing mapping routines. Thanks to P. Teenjes and M. Einasto for help. EB acknowledges the financial support provided by MIUR PRIN 2011 ``The dark Universe and the cosmic evolution of baryons: from current surveys to Euclid'' and  Agenzia Spaziale Italiana for financial support from the agreement ASI/INAF/I/023/12/0. We thank S.~Borgani and G.~Murante for giving us access to the simulations analysed in this paper. CG thanks CNES for financial support.
\end{acknowledgements}

\bibliographystyle{aa} 
\bibliography{whimbib} 

\begin{appendix}

\section{HOD Formalism}
\label{app:a}

We applied halo occupation distribution (HOD) formalism to our adopted large-scale simulation of \citet{2012MNRAS.423.2279C} in order to construct the galaxy distribution. Dark matter halos and subhalos represent tracers of the central galaxies and satellites, respectively \citep{2000MNRAS.318.1144P, 2004ApJ...609...35K, 2007MNRAS.376..841V}. We thus started by producing a DM halo catalogue by applying the Friend-of-Friend (FoF) algorithm with linking length parameter $b=0.2$ (in the units of the mean inter-particle separation) to the simulated data. This choice of the linking length value yields mass functions consistent with different theoretical predictions of the virial mass function \citep[e.g.][]{2004MNRAS.355..819G, 2005MNRAS.364.1105S}. The resulting FoF catalogue contains the position and the virial mass of each DM halo. In the following we describe how we used the FoF catalogue to obtain the magnitudes and positions of the galaxies.

\begin{figure}
\centering
\includegraphics[width=8cm]{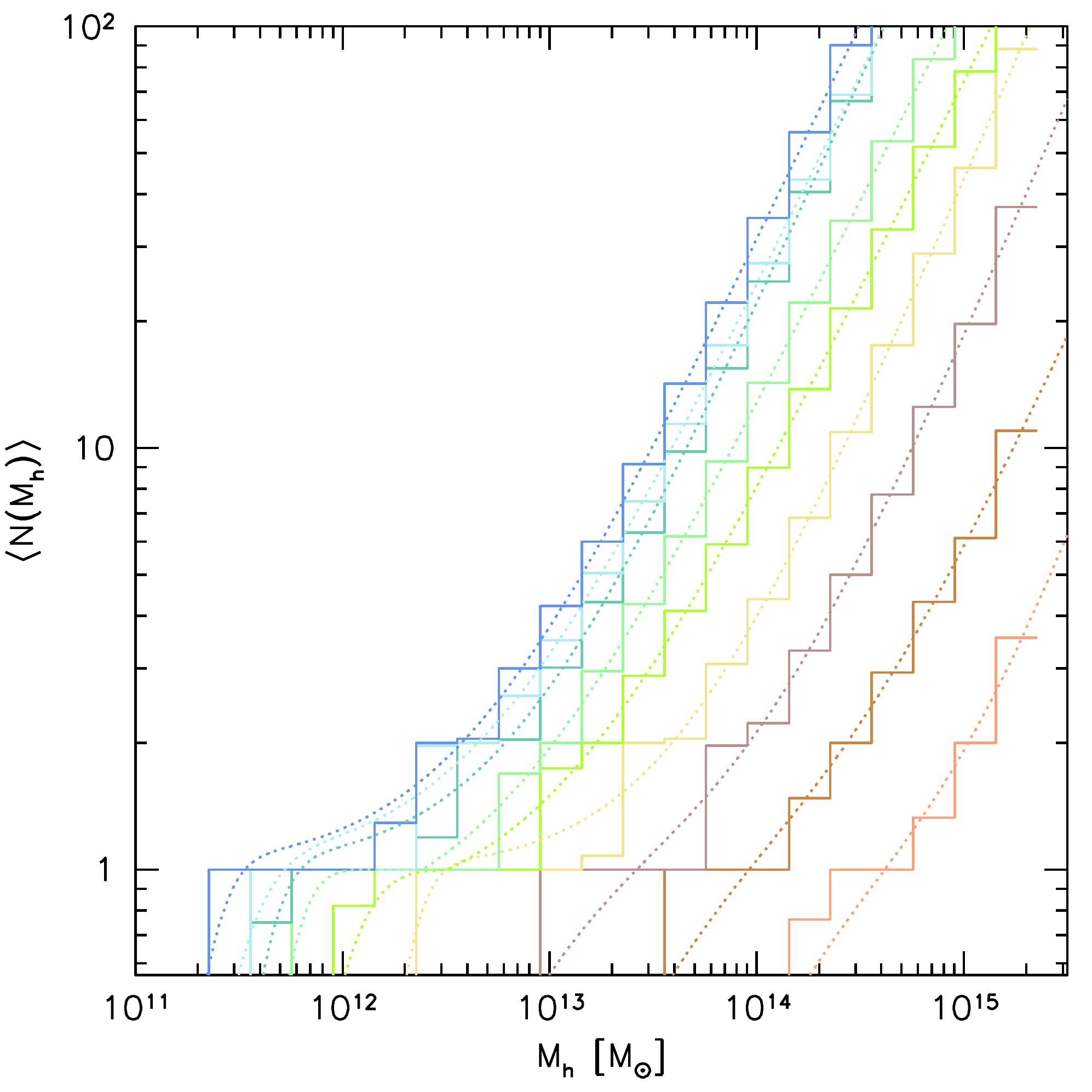}
\caption{The halo occupation number, i.e. the mean number of galaxies in a halo of a given virial mass, is shown in different luminosity bins (colour coding as in Fig.~\ref{figDHOD}). The solid histograms represent the statistical realisations of the galaxies that populate the halos extracted from the \citepalias{2012MNRAS.423.2279C} simulation and their substructures. The dotted curve shows the predictions from Eq.~\ref{eqhod} using the parameters in \citet{2011ApJ...736...59Z}.}
\label{fHOD}
\end{figure}

\begin{figure*}
\centering
\includegraphics[width=18cm]{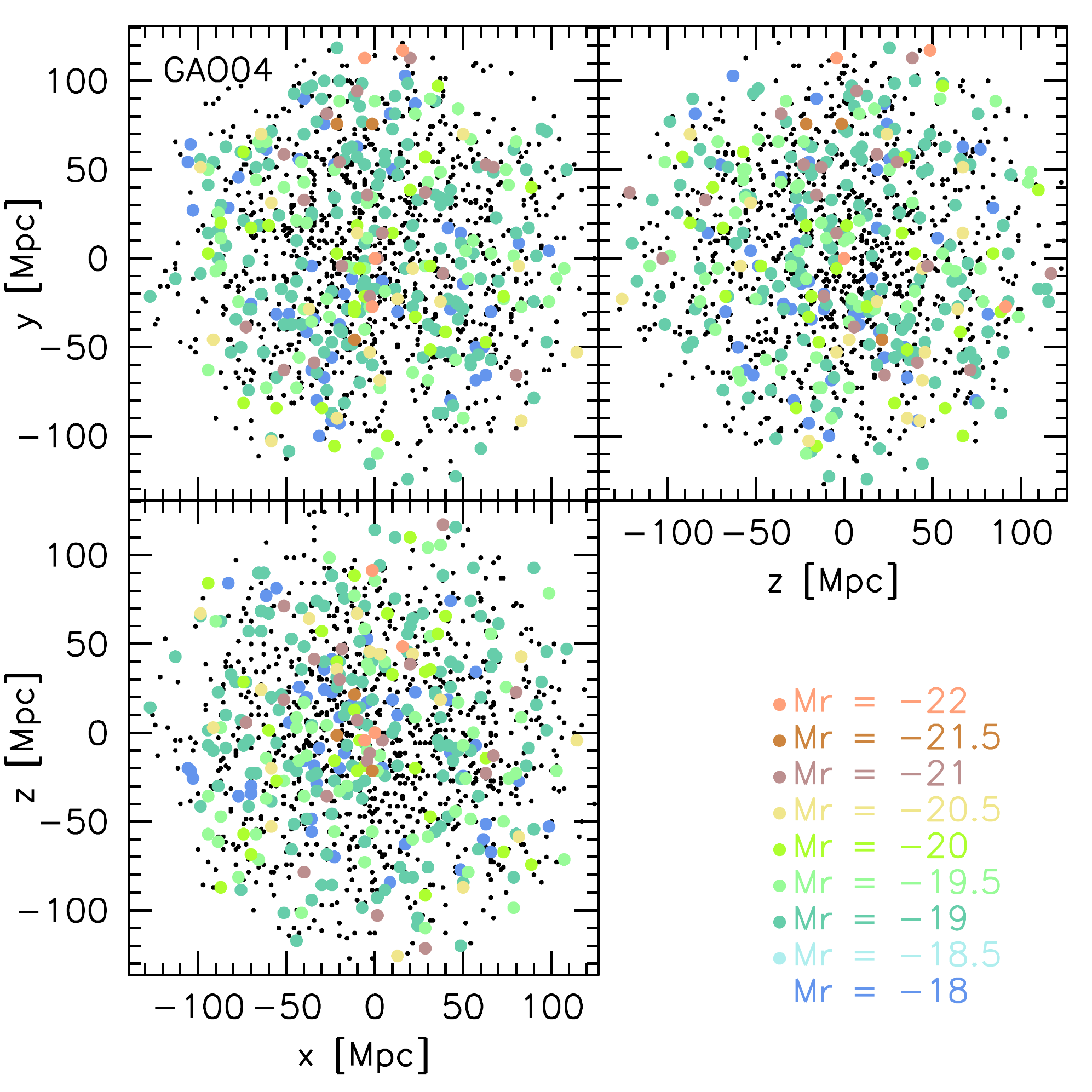}
\caption{Three orthogonal projections of the distribution of satellite galaxies in centres of DM halos of $ \sim$1~Mpc radius in our adopted simulation of 
\citetalias{2012MNRAS.423.2279C} (coloured dots). The colour coding indicates the magnitude of a given galaxy. The black dots show the positions of galaxies fainter than M$_r$=-18 that populate halo and sub-halos in the simulation according to the mass function by \citet{2010MNRAS.404..502G} down to $10^{10}~M_{\odot}$.}
\label{figDHOD}
\end{figure*}

\begin{enumerate}
\item
For each DM halo we used its FoF mass from above to calculate the virial radius $R_\mathrm{vir}$, to know up to which scale to populate it with subhalos. This was done according to the spherical collapse formalism which yields the estimate of the virial overdensity $\Delta_\mathrm{vir}(z)$ \citep[e.g.][]{1980lssu.book.....P, 1996MNRAS.282..263E, 1996MNRAS.280..638K, 1998ApJ...495...80B} which is linked to the virial mass and the virial radius as:
\begin{equation}
M_h = \frac{4 \pi}{3} R_\mathrm{vir}^3 \frac{\Delta_\mathrm{vir}(z)}{\Omega_\mathrm{m}(z)}
\Omega_\mathrm{m}(0) \rho_c\,
\end{equation}
where $\rho_c(z)$ and $\Omega_\mathrm{m}(z)$ represent the critical density and the cosmological matter density parameter at redshift $z$, respectively.

\item
We assigned each halo with a concentration parameter according to mass-concentration relation of \citet{2001MNRAS.321..559B}, assuming a log-normal scatter 
$\sigma_{\ln c} = 0.25$ around the mean value.

\item
We populated each halo with subhalos by performing Monte-Carlo realisations of the subhalo mass function model of \citet{2010MNRAS.404..502G} which features both a  
redshift evolution and a concentration dependence of the subhalo mass function. The model assumes that the spatial distribution of the subhalos is less
concentrated than the NFW DM profile \citep{1996ApJ...462..563N}, since this includes the effects of the dynamical friction and the tidal stripping \citep{2004MNRAS.355..819G, 2004MNRAS.352.1302V, 2008MNRAS.386.2135G}.
Thus, for  each halo we  now have  the subhalo populations with known positions and  masses.

\item
We then assigned each halo and subhalo with a galaxy with a luminosity value according to the abundance matching approach \citep[see e.g.][]{2010ApJ...717..379B} as follows. In the core of the method is the halo occupation function
 \begin{eqnarray}
  \langle N(M_h) \rangle = \frac{1}{2} \left[ 1 + \mathrm{erf} \left( \frac{\log M_h - \log M_\mathrm{min}}{\sigma_{\log M}} \right) \right] \nonumber \\
  \times \left[ 1 + \left(  \frac{M_h - M_0}{M'_1}\right)^{\alpha} \right]   , \label{eqhod}
  \end{eqnarray}
which describes the mean number of halos within a parent halo of mass $M_h$ \citep{2011ApJ...736...59Z}.

\begin{figure}
	\centering
\vspace{-1cm}
\includegraphics[width=9.0cm,angle=0]{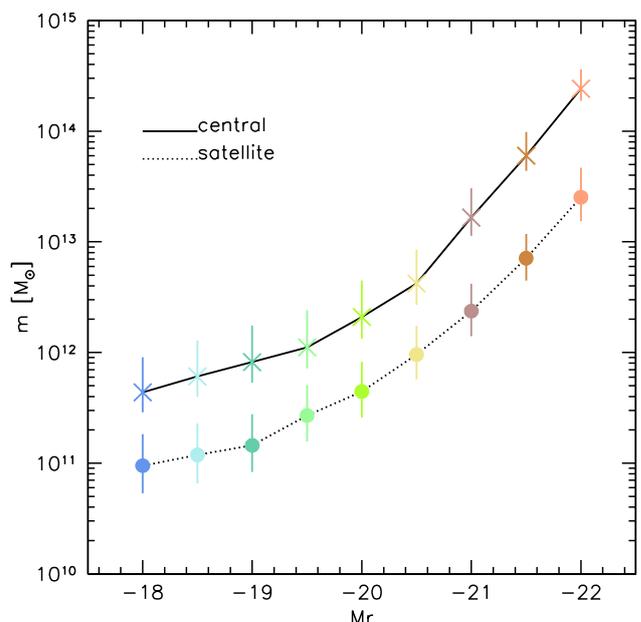}
\vspace{-3cm}
\caption{Median masses of parent halos (solid line and crosses) and subhalos (dotted line and dots) hosting galaxies with different r band absolute magnitudes, when applying the HOD formalism to \citetalias{2012MNRAS.423.2279C} simulations (colour coding as in Fig.~\ref{figDHOD}).}
\label{massmedian.fig}
\end{figure}

We then made the standard assumptions that 1) all subhalos are populated by one galaxy only, i.e. that the number of substructures into which we have resolved the parent halos is sufficient to host at most one galaxy and that 2) there are no ``orphan'' subhalos. The latter hypothesis is justified by the fact that there is no evidence for massive dark halos with no baryon content. Using these assumptions Eq.~\ref{eqhod} then describes the mean number of {\it galaxies} within a parent halo of mass $M_h$.

The values of the parameters of Eq.~\ref{eqhod} have been found by \citet{2011ApJ...736...59Z} by fitting the projected SDSS-DR7 2-point galaxy-galaxy correlation function of \citet{2009ApJS..182..543A} in the luminosity range $M_r=-[18.5,-22]$ sampled in seven, equally spaced bins. Thus, the parameters of Eq.~\ref{eqhod} are different for the different luminosity bins.

We then applied the method to our FoF DM halos obtained from the \citetalias{2012MNRAS.423.2279C} simulations as follows: for a given parent halo of mass $M_h$ we use Eq.~\ref{eqhod} to determine the mean number of galaxies at a given magnitude $M_r$ (see Figs.~\ref{fHOD} and~\ref{massmedian.fig}). We repeated the procedure for each magnitude bin from $M_r=-18.5$ to $M_r=-22$ thus obtaining the luminosity function N($L_r$) for a given host halo.

We then ordered the above galaxies from most luminous to least luminous. For the same parent halo we went back to the subhalo distribution obtained above \citep{2010MNRAS.404..502G} and ordered the subhalos from the most massive to the least massive. We then matched the parent halo or its subhalo of a given mass ranking by a galaxy with the same luminosity ranking (most massive with the most luminous etc.) until all galaxies were assigned to the subhalos. We removed the low mass subhalos which were assigned with no galaxies.

The outcome is a set of parent halos, extracted from the C21 simulation, containing a set of subhalos.
Each halo and subhalo has a galaxy at its centre with known location and $r-$band luminosity (see Fig.~\ref{figDHOD}).

\end{enumerate}

\end{appendix}

\end{document}